\documentclass[useAMS,usenatbib,fleqn,usedcolumn]{mnras}

\usepackage[british]{babel}             
\usepackage[T1]{fontenc}                
\usepackage{graphicx}                   

\usepackage{bm}
\usepackage{amsmath}
\usepackage{amssymb}

\renewcommand{\vec}[1]{\ensuremath{\bm{{#1}}}} 
 
\newcommand{\unit}[1]{\ensuremath{\,\mathrm{#1}}}
\newcommand{\diff}{\ensuremath{\mathrm{d}}}
\newcommand{\deriv}[2]{\ensuremath{\frac{\diff #1}{\diff #2}}}
\newcommand{\tderiv}[2]{\ensuremath{\frac{\diff #1}{\diff #2}}}
\newcommand{\St}{\ensuremath{\mathit{St}}}
\newcommand{\delv}{\ensuremath{\delta\!v}}
\newcommand{\delr}{\ensuremath{\delta\!r}}
\newcommand{\vmed}{\ensuremath{\Delta v_{\rm med}}}

\title[Collisions in self-gravitating discs ]{Collision velocity of dust grains in self-gravitating protoplanetary discs}
\author[Booth \& Clarke]{Richard A. Booth\thanks{E-mail: rab200@ast.cam.ac.uk}$^1$, Cathie J. Clarke$^1$\\
$^{1}$Institute of Astronomy, University of Cambridge, Madingley Road, Cambridge, CB3 0HA, United Kingdom}

\begin{document}

\date{Accepted 2016 February 25. Received 2016 February 25; in original form 2015 November 28}

\pagerange{\pageref{firstpage}--\pageref{lastpage}} \pubyear{2015}

\maketitle

\label{firstpage}

\begin{abstract}
We have conducted the first comprehensive numerical investigation of the relative velocity distribution of dust particles in \emph{self-gravitating} protoplanetary discs with a view to assessing the viability of planetesimal formation via direct collapse in such environments. The viability depends crucially on the large sizes that are preferentially collected in pressure maxima produced by transient spiral features (Stokes numbers, $\St \sim 1$); growth to these size scales requires that collision velocities remain low enough that grain growth is not reversed by fragmentation. We show that, for a single sized dust population, velocity driving by the disc's gravitational perturbations is only effective for $St > 3$, while coupling to the gas velocity dominates otherwise. We develop a criterion for understanding this result in terms of the stopping distance being of order the disc scale height. Nevertheless, the relative velocities induced by differential radial drift in multi-sized dust populations are too high to allow the growth of silicate dust particles beyond $\St \sim 10^{-2}$ or $10^{-1}$ ($10\unit{cm}$ to m sizes at $30\unit{au}$), such Stokes numbers being insufficient to allow concentration of solids in spiral features. However, for icy solids (which may survive collisions up to several $10\unit{m\,s}^{-1}$), growth to $\St \sim 1$ ($10\unit{m}$ size) may be possible beyond $30\unit{au}$ from the star. Such objects would be concentrated in spiral features and could potentially produce larger icy planetesimals/comets by gravitational collapse. These planetesimals would acquire moderate eccentricities and remain unmodified over the remaining lifetime of the disc.  
\end{abstract}


\section{Introduction}
\label{Sec:Intro}

While spectral modeling of protoplanetary discs indicates that growth of regions up to mm or cm sizes \citep{Ricci2010}, it has been argued based on models of {\it non-}self-gravitating discs that agglomerative growth beyond these sizes becomes inefficient due to bouncing, fragmentation and radial drift \citep[e.g.][]{Brauer2008,Zsom2010,Pinilla2012,Garaud2013}. How growth proceeds beyond cm sizes remains one of the biggest uncertainties in current models of planet formation.

Collective phenomena, such as the streaming instability \citep{Youdin2005,Johansen2007}, may provide a possible growth mechanism, but typically this requires large dust-to-gas ratios to be effective \citep{Bai2010b}, potentially limiting the process until late phases of disc evolution. Alternatively, in young discs massive enough to be `gravoturbulent'\footnote{i.e. in a self-regulated state where cooling is balanced by spiral shock heating, thus maintaining a Toomre $Q$ parameter close to unity.}, the spiral structure may be able to focus large grains strongly enough that planetesimals may form directly via gravitational collapse in the dust layer \citep{Rice2004,Rice2006,Gibbons2012,Gibbons2014} and thus provides a mechanism by which planetesimals may be formed early within the disc's lifetime.

The focussing of dust grains requires Stokes numbers approaching unity, where the Stokes number, $\St$, is the ratio of the stopping time to dynamical time-scale. For discs massive enough to be self-gravitating, $M_{\rm D} \sim 0.1 M_{\sun}$, this requires growth to sizes of several metres at 30 AU. While growth to $\St\sim 1$ may occur in less massive discs (and where $\St \sim 1$ corresponds to smaller objects, roughly cm-sizes at 30\unit{au} in a minimum mass solar nebular disc), gravoturbulence gives rise to perturbations in the velocity of order the sound speed in both the gas and dust \citep{Gibbons2012,Walmswell2013}. Collisions at the sound speed would inevitably lead to fragmentation, since the sound speed at $30\unit{au}$ is a few $100\unit{m\,s}^{-1}$, well above the typical fragmentation threshold for both silicate ($\sim 1\unit{m\,s}^{-1}$, \citealt{Guttler2010}) and icy grains (a few $10\unit{m\,s}^{-1}$, \citealt{Wada2009,Gundlach2015}). However, collision velocities can be considerably lower than the global velocity variation as particle pairs at small separations will experience similar perturbations, resulting in only small changes to their relative velocity. Clearly, in order that planetesimals can form via direct gravitational collapse of a layer of solid material, the collision velocity has to remain low enough for particles to first grow to metre sizes and beyond in a low velocity dispersion environment.

Observationally, the detection of mm to cm-sized grains in Class I discs \citep{Greaves2008,Miotello2014} has shown that grains can indeed grow to relatively large sizes while the discs remain massive. Typical mass estimates for Class I discs are in the range $\sim 0.01M_\odot$, up to $\sim 0.1\,M_\odot$, modulo the typical uncertainty in the dust-to-gas ratio and grain opacity \citep{Eisner2012}, with younger systems ($\lesssim 1\unit{Myr}$) perhaps favouring the higher masses conducive to strong self-gravity \citep{Mann2015}. Since many of these systems have large radii (several $100\unit{au}$), the large mass alone does not necessarily imply that self-gravity is important; however, some may well be in the self-gravitating regime (e.g. WL 12, \citealt{Miotello2014}). Furthermore, it is possible that grain growth to $\sim 10\unit{cm}$ sizes may have already occured in HL Tau \citep{Zhang2015}, a very young system that may have recently been in the self-gravitating phase, even if the lack of spiral structure in recent high resolution observations imply it may no longer be so \citep{ALMA2015}. The formation of such large pebbles so early in the disc's evolution presents exciting prospects for theories such as the planet formation via the streaming instability \citep{Youdin2005,Johansen2007} or pebble accretion \citep{Lambrechts2012} and perhaps even dust driven gravitational fragmentation.

In contrast to gravoturbulent discs, the relative velocity of dust particles in MRI and classical turbulence has been extensively studied with both theoretical and numerical approaches. In an astrophysical context, the classical picture is that eddies with turnover times shorter than the stopping time of the particles give rise to random kicks that affect the relative velocity while those with larger turnover times lead to correlated motion \citep{Volk1980,Ormel2007}. Additionally, \citet{Pan2010} showed that the size of the eddies is also important, since eddies larger than the separation between two particles will cause the particles to experience similar kicks. Thus the history of the separation between particle pairs is important for determining the collision velocity. Numerical simulations have been used to test these  ideas, showing good agreement with the theory \citep{Cuzzi2003, Bai2010a, Carballido2010, Pan2010, Pan2013, Pan2014a}. The relative velocity of dust particles in self-gravitating discs has received much less investigation, although \citet{Gibbons2012} investigated the distribution of individual particle velocities in a shearing box. For MRI turbulent discs, the gravitational acceleration due to MRI density fluctuations has also been found to be important for large bodies with $\St \gg 1$ \citep{Laughlin2004,Ida2008}.

In this work we first present a qualitative picture for the dynamics of dust particles in self-gravitating discs and describe how this controls the velocity of collisions between pairs of particles, focussing primarily on particles with $\St \gtrsim 0.3$: i.e. those that can be resolved in simulations. We then use smoothed particle hydrodynamics simulations (SPH) to quantify the collision velocities. We then use our results to address the prospects for in situ grain growth and the possibility of planetesimal formation by gravitational collapse in spiral arms.

In section \ref{Sec:CollVell} we discuss the important effects that govern the motion of dust particles in self-gravitating discs. In sections \ref{Sec:Method} and \ref{Sec:IC} we describe the simulations. Sections \ref{Sec:DustDisp} and \ref{Sec:Sample} discuss how we measure the collision velocity and the associated biases. We discuss the collision velocity and concentration of identical particles in section \ref{Sec:MonoDisp}, while in section \ref{Sec:BiDisp} the case of differing particle sizes is considered. Finally in sections \ref{Sec:Discuss} and \ref{Sec:Conclude} we present our discussion and conclusions.

\section{Collision velocities in self-gravitating discs}
\label{Sec:CollVell}

In order to interpret the simulations presented in this work, it is useful to first consider the factors that affect the velocity of dust particles in self-gravitating discs. For simplicity, throughout this section we focus on the case of identical particles (mono-disperse case), unless otherwise stated. The case  of differing particle sizes (bi-disperse case) is discussed in detail in \autoref{Sec:BiDisp}. Our discussion contains a number of similarities with the classical turbulent case -- particularly that of \citet{Pan2010} and \citet{Pan2013}, in that a comparison of length scales and particle separations is central to our argument.

For large particles that are weakly coupled to the gas, i.e. those with $\St \gg 1$, the dynamics of a single particle is well described by a series of perturbations through gravitational scattering by the spiral structure \citep{Walmswell2013}. For a disc with mass 10 per cent of the star mass and a short cooling time -- 5 times the dynamical time-scale (which represents a region just inside the point at which the amplitude of spiral features becomes sufficient to drive gravitational fragmentation in the gas) -- the scattering events were able to drive the eccentricity, $e$, of planetesimals to typical values of $e \sim 0.1$\footnote{\citet{Walmswell2013} neglect drag forces entirely, but eccentricity growth to $e \sim 0.1$ is consistent with $\St > 100$ for their simulation parameters.}. This eccentricity in combination with random phases leads to the crossing of orbits \citep[c.f. caustics for turbulent gases,][]{Gustavsson2011} which dominates the collision velocity. 

For initial conditions that are dynamically cold, with particles at similar positions having similar velocities, then correlations in the motions affect the distributions of eccentricities and relative velocities. Over time these correlations are broken by variations in the gravitational forces on scales of order the particle separation. However, since there is little power on these scales \citep{Boley2007,Cossins2009,Michael2012}, the correlation times can be long.

 Using 3D simulations \citet{Boley2007} reported that the strength of the density perturbations follow the relation $\delta \rho \sim (m^2 + m_0^2)^{-k}$, with $m_0 \sim 13$, and $k=1.5$ where $m$ is the azimuthal wavenumber. Our 2D simulations show good agreement,  but are better fit with $m_0=16$ and $k=1.2$ to $1.4$,  showing a slightly slower decrease in power to small scales. The length scale, $\lambda$, associated with a given mode is $\lambda/R \sim 1/m$, thus $m_0 = 16$ corresponds to $\lambda/R \approx 0.06 \approx 2 H/R$, showing that the power is predominantly on scales $\gtrsim H$. The rapid decrease in power towards small scales forms the basis for our argument that particles with small separations will only experience small differences in the gravitational perturbations and that the disc scale height is the natural length scale to associate with this process. However, even though the difference in accelerations between two closely separated particles is small, if the drag forces are sufficiently weak the perturbations may eventually break the correlations and generate large relative velocities and significant eccentricities, as found by \citet{Walmswell2013}.

For smaller particles, for which the drag force is non-negligible, the damping provided by drag forces can prevent the growth of large relative velocities for pairs with small separations. Since the damping occurs on a stopping time, perturbations growing on time-scales longer than this are damped and will not contribute significantly to the relative velocity. For a collision occurring at velocity $\Delta v$, we can define a characteristic distance $\lambda_{\rm stop}$, which is the typical distance that the particles cross within a stopping time prior to the collision. Perturbations that vary on scales much larger than this will have similar effects on the velocity of both particles, but only a weak effect on their relative velocity. Therefore, it is natural to compare the driving scale of the perturbations, $\lambda_{\rm d}$, to the characteristic stopping distances of colliding particles
\begin{equation}
\lambda_{\rm stop} = \Delta v t_s,
\end{equation}
where $\Delta v$ is the typical relative velocity.  By equating $\lambda_{\rm d}$ and $\lambda_{\rm stop}$, we can determine a critical relative velocity,
\begin{equation}
 v_{\rm crit} = \frac{\lambda_{\rm d}}{H} \St^{-1} c_s. \label{Eqn:v_crit}
\end{equation}
Since there is little power in the density field on scales $<H$, we expect that once the typical collision velocity is smaller than $\St^{-1} c_s$ driving by self-gravity should become inefficient. If only driving by self-gravity is possible then decreasing $\St$ further should lead to a rapid drop in the typical collision velocity. However, as $\St$ decreases and the drag forces increases, the coupling of the dust velocity to the gas velocity becomes stronger and driving by fluctuations in the gas velocity on scales $ \le \lambda_{\rm stop}$ becomes more important. In \autoref{Sec:MonoDisp} we demonstrate the transition from gravitational driving to driving by drag and find the corresponding Stokes number at which this occurs.

The bi-disperse case is further complicated by the differing terminal velocity of particles with different stopping times, $v_{\rm term} = t_s \nabla P / \rho $. The different terminal velocities result in different azimuthal and radial drift velocities even for axisymmetric discs. For small particles ($\St \ll 1$) with very different stopping times this could give rise to a mean collision velocity that is larger than the typical velocity dispersion within a given size bin. For large particles ($\St \gg 1$), the relative radial and azimuthal drift velocity is  small, but the velocity dispersion is large (due to efficient driving). These behaviours are tested in \autoref{Sec:BiDisp}.

\section{Numerical Methods} 
\label{Sec:Method}
We have run two-dimensional smoothed particle hydrodynamics simulations of self-gravitating discs using a modified version of the \textsc{gadget-2} code \citep{Springel2005}. Below we briefly describe the improvements made to the hydrodynamics and the implementation of dust dynamics.

\emph{Kernel} -- 
We use  the Wendland $C^6$ kernel, which is stable against the pairing instability and improves convergence in the presence of strong shear \citep{Dehnen2012}. We define the smoothing length, $h$ as the full extent of the kernel, as in \citet{Springel2005}. We set the smoothing lengths to give a constant number of neighbours in two dimensions,
\begin{equation}
\pi h_i^2 \sum_j W(|\vec{r}_{ij}|, h_i) = N_{\rm ngb}, \label{Eqn:Smooth}
\end{equation}
where $W$ is the kernel function, $\vec{r}_{ij} = \vec{r}_i - \vec{r}_j$, $h_i$ is the smoothing length of the $i$th particle and $N_{\rm ngb}$ is the desired number of neighbours. Typically we use $N_{\rm ngb} = 50$ or $100$ to reduce noise and ensure an accurate density estimate.\footnote{$N_{\rm ngb} = 50$, corresponds to $\eta = 4$, for $\eta$ as defined in \citealt{Price2012}.}

\emph{Gradients} -- 
Additionally we make use of the integral approximation to kernel gradients
\citep{Garcia-Senz2012,Rosswog2015}. The gradient of a quantity $f$ is given by
\begin{equation}
 \nabla^k f(\vec{r}) = \sum_j \frac{m_j}{\rho_j} f_j 
	\sum_{d=1}^2 C^{kd}(\vec{r}, h) (r^d-r^d_j) W(|\vec{r} - \vec{r}_j|, h), \label{Eqn:IGrad}
\end{equation}
where the upper indices denote the spatial dimensions and the lower indices denote the particles. The matrix ${C}^{kd} = ({\tau}^{-1})^{kd}$, where 
\begin{equation}
 \tau^{kd}(r_i, h_i) = \sum_j \frac{m_j}{\rho_j} {r}^k_{ij} {r}^d_{ij} W(r_{ij}, h).
\end{equation}
The quantities inside the second sum in \autoref{Eqn:IGrad} replace directly the $\nabla W$ terms that appear in the equations of motion. SPH formulations based on the integral approximation produce less numerical noise, better resolve instabilities and are not sensitive to the underlying particle distribution, thus preserving symmetry more accurately. For details see \citet{Garcia-Senz2012} or \citet{Rosswog2015}. Similarly, in \citet{Booth2015} we found integral gradients introduce less numerical noise into the dust particle velocities.

\emph{Viscosity} -- 
To ensure that artificial dissipation does not interfere with the angular momentum transport by gravitational torques and Reynolds stress we use a \citet{Cullen2010} type viscosity switch to reduce viscosity away from shocks. Additionally we employ the noise trigger of \citet{Rosswog2015} to help maintain particle order in the presence of strong shear. Since the shocks in self-gravitating discs are typically weak, we use a higher than normal value of the maximum viscosity parameter, $\alpha_{\rm max} = 3$ to ensure sufficient entropy generation at shock fronts. Away from shocks the scheme produces $\alpha  < 0.1$, i.e. less viscosity than is often used with fixed $\alpha$ prescriptions \citep{Lodato2004,Rice2004,Rice2006,Meru2011,Meru2012}.

\emph{Conductivity} --
We employ artificial conductivity as in \citet{Price2012} to smooth jumps at contact discontinuities.

\emph{Self-gravity} --
For self-gravity we use a \citet{Barnes1986} tree with a relative opening criterion, as described in \citet{Springel2005}. We soften the gravitational force on scales smaller than $H = c_s / \Omega$ to take into account the fact that we are modelling a 3D dimensional structure in 2D. This is necessary since otherwise the gravitational force on scales smaller than $H$ is over-estimated, which leads to the formation of artificial small scale structures and can lead to spurious fragmentation \citep{Muller2012,Young2015}. Following \citet{Muller2012}, we soften the force by integrating the force felt by each particle over the vertical structure of the disc, which we approximate by a Gaussian. For full details, see Appendix \ref{Sec:Gravity}.

\emph{Dust Dynamics} -- 
We include dust particles in the test-particle limit, which feel gravitational and drag forces, but do not affect the motion of the gas. The dust acceleration is given by
\begin{equation}
 \deriv{\vec{v}_d}{t} = - \frac{\vec{v}_d - \vec{v}_g}{t_s} + \vec{g},
\end{equation}
where $\vec{v}_d, \vec{v}_g$ are the dust and gas velocities, $t_s$ is the stopping time, and $\vec{g}$ if the gravitational acceleration. As in \citet{Booth2015}, we use a semi-implicit update for the dust velocity as part of the kick-drift-kick scheme in \textsc{gadget-2}, where the velocity is set using the analytical solution for the drag force, under the assumption that the gas velocity, gravitational acceleration and stopping time remain constant throughout the time step. These quantities are calculated by interpolating the gas properties at the location of the dust particle, with the \emph{dust} smoothing length calculated via \autoref{Eqn:Smooth}, summing over neighbouring \emph{gas} particles to ensure sufficient neighbours.  The robustness of the scheme has been demonstrated in a wide range of idealized test problems \citet{Booth2015}.

For the drag force, we set the stopping time using either a constant Stokes number, $t_s = \St \Omega^{-1}$, or a scale-free approximation to the Epstein regime. In the Epstein regime the stopping time for spherical grains is $t_s = \rho_s s / (\rho_g v_T)$, where $\rho_s$, $s$, $\rho_g$ and $v_T = \sqrt{\pi/8} c_s$ are the internal density of the grain, the grain size, the gas density and the mean thermal velocity of the gas respectively. Since in the mid-plane, $\rho_g \sim \Sigma / H$ and $H = c_s / \Omega$, where $\Sigma$ is the surface density and $c_s$ is the sound speed, we see that $\St = t_s \Omega \propto s / \Sigma$, with no dependence on the sound speed. Therefore we can include the drag force in a scale free way by taking $t_s = \St \Sigma_0 / (\Sigma \Omega)$, where $\Sigma_0$ is the the surface density in the absence of gravitational perturbations.

We have neglected the effects of collisions between dust grains on the dust dynamics, since for typical parameters the time between collisions, $t_c$, is much longer than the stopping time, $t_s$,
\begin{equation}
 \frac{t_c}{t_s} = (n \sigma \langle v \rangle)^{-1} \frac{\rho_g v_T}{\rho_s s}
 = \frac{4}{3} \sqrt{\frac{8}{\pi}} \frac{\rho_g c_s}{\rho_d \langle v \rangle}
\approx 2.13 \frac{\rho_g c_s}{\rho_d \langle v \rangle}, \label{Eqn:CollTime}
\end{equation}
where $\rho_d$ is the local mass density of the dust and $\langle v \rangle$ is the typical collision velocity. For typical dust-to-gas ratios of 0.01 and collision velocities of order $c_s$ or smaller, $t_c/t_s > 200$.

\emph{Cooling} -- We assume a $\beta$ cooling law, $\dot{u} = - u / t_c$, where ${t_c = \beta / \Omega}$. We consider $\beta$ in the range $7 \le \beta \le 25$, to investigate the dependence of the collision velocity on the strength of the gravitational perturbation, while remaining safely in the range in which fragmentation does not occur in 2D \citep{Young2015} as well as minimizing the influence of artificial
viscosity heating on the spiral structure.

\section{Disc model and initial conditions}
\label{Sec:IC}

\begin{figure}
 \centering
\includegraphics[width=\columnwidth]{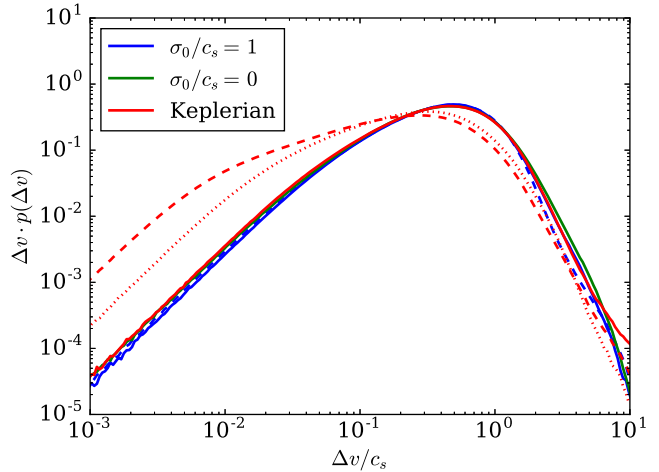}
\caption{Collision velocity distribution after 200 inner dynamical time-scales for $\St = 10$ and different initial conditions at resolutions of $10^{6}$ (solid), $4 \times 10^6$ (dashed) particles per phase. The dotted line shows the high-resolution simulation sub-sampled at the same resolution of the low resolution simulation. The line colour refers to the type of initial conditions used. Blue and green lines refer to simulations that used the gas velocity as the base for the initial dust velocity, with differing amounts of noise introduced into the velocity. The red lines refer to simulations started with initially Keplerian velocities. 
}
\label{Fig:VelPDF_IC}
\end{figure}

\begin{figure*}
 \centering
\includegraphics[width=\textwidth]{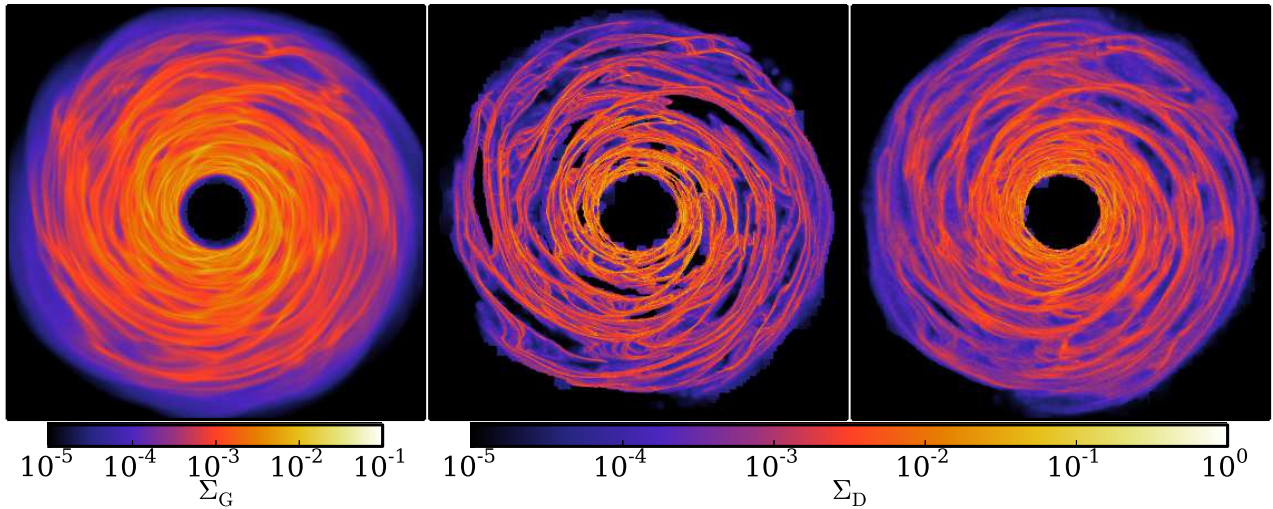}
\caption{\emph{Left}: Gas surface density. \emph{Middle \& Right}: Dust surface density for $\St = 3$ and $\St = 10$ respectively. The results are for a simulation with $4 \times 10^6$ particles per phase and cooling parameter $\beta = 10$, shown 200 inner dynamical time-scales ($\Omega^{-1}$) after the dust was introduced. The spatial range covers the whole disc from $-5$ to $5$ and the dust density is scaled to give the same background density as the gas. The overall dust morphology is similar to the gas in both cases, but with narrower spiral features for $St \sim 1$. The majority of dust particles are found in regions where the dust density is enhanced relative to the average $\Sigma_{\rm D} / \Sigma_{\rm G}$. 
}
\label{Fig:DiscDens}
\end{figure*}

For the gas disc model we take a central mass $M_\star = 1$ and disc mass $M_D = 0.1$. For the disc structure we use an initial surface density profile $\Sigma \propto R^{-2}$, between $R_{\rm in} = 1$ and $R_{\rm out} = 5$, which ensures the same resolution in $h / H$ everywhere. We set up the disc with an initial Toomre parameter $Q = 2$, and evolve it for $10^3$ inner dynamical times, approximately nine cooling times at the outer edge, allowing the self-gravitating structure to reach a steady state before introducing the dust.

The dust particles are given an initially uniform density with the same $\Sigma \propto R^{-2}$ density distribution as the gas. The simulations have then been run for a further $100$ to $300$ inner dynamical times, until the dust velocity distribution has stopped varying. For our canonical simulations we use $10^6$ particles per phase. We use the same number of gas neighbours for both the gas and dust particles, $N_{\rm NGB} = 50$, which corresponds to $h/H \approx 0.2$. It is worth noting that our chosen kernel, the Wendland C$^6$, is 1.35 times wider than the commonly used Cubic Spline kernel, such that sound waves with wavelength $\lambda \gtrsim 0.6 H$ should be well resolved \citep{Dehnen2012}. Additionally we have run some high-resolution simulations using $4 \times 10^{6}$ and $16 \times 10^{6}$ particles per phase, in which we use a higher $N_{\rm NGB} = 100$ to further reduce noise. 

Since the velocities of particles with small initial phase space separations remain correlated for many orbits, care needs to be taken to ensure the initial conditions do not affect the measured relative velocity distribution. This correlation time will inevitably be resolution dependent since small wavelength perturbations are responsible for breaking these correlations. Rather than trying to model the processes responsible (e.g. gravitational interactions between planetesimals for $\St \gg 1$), we calculate the steady-state velocity distribution, once the initial correlations have been lost. In doing so we neglect a possible low velocity component of relative velocity distribution that may play a role in the growth of grains, but are able to determine the frequency of high velocity collisions that will be responsible for fragmentation. However, since the formation process itself may take several dynamical times, it is likely that highly correlated initial conditions are not representative of the distribution of velocities that the dust is initially formed with. Similar non-convergence at the smallest resolvable scales is seen in turbulence driven by the streaminng instability \citep{Carrera2015}.

We therefore choose our initial conditions in such a way that the steady state distribution of dust velocities is reached as rapidly as possible. Since the particle pairs that take the longest to reach steady state are those with small separations, it is sensible to break the initial correlation of particles that are initially close together. The simplest way to achieve this is to add noise to the initial particle velocity, thus ensuring that particles close together in position space have a significant separation in the full phase space. For our standard set-up, we set the initial dust velocity to the local gas velocity, on top of which we add a random velocity of order the sound speed. In adding an initial velocity dispersion we now have to integrate for long enough to ensure that the measured velocity distribution is not just an artefact of the initial velocity distribution. For tightly coupled particles integrating for relatively few dynamical times is sufficient (though in practice we do not use less than 100 inner dynamical times). For the more weakly coupled particles, the velocity dispersion grows above the initial velocity dispersion within a few 10s of dynamical time-scales, therefore it is sufficient to integrate until a steady state is reached and the velocity dispersion saturates.

The effects of the initial conditions on the distribution of relative velocities are demonstrated in \autoref{Fig:VelPDF_IC}, which shows the relative velocities for simulations at resolutions of $1 \times 10^6$ and $4 \times 10^6$ particles per phase and $\St = 10$ (for an explanation of how the velocity dispersion is calculated, see \autoref{Sec:DustDisp}.). At $1 \times 10^6$ particles per phase we find that the velocity correlations are lost before 200 inner dynamical times, resulting in the same distributions independent of whether noise was added, or whether the initial velocity was set to the gas velocity or the local Keplerian velocity. However, for $4 \times 10^6$ this is no longer the case and the simulations with correlated initial conditions produce distributions biased to lower velocities. 

By sub-sampling the dust in the $4 \times 10^6$ particle simulation to the same resolution as the $10^6$ particle simulation we see that the effect is not merely due to sampling the distribution on smaller scales, since it only partially accounts for the difference, but that the correlation time is also longer.  This is because the high resolution simulations have \emph{less} power on small scales due to reduced noise, which means the correlation time is longer at small separations. It is important to note that the although we measure the velocity dispersion on these small scales, the velocity dispersion is driven by effects on larger scales ($\lambda_{\rm d} > \lambda_{\rm sam}$, see \autoref{Sec:Sample}) so the steady state velocity dispersion is correctly reproduced. However, the correlation times at small separations are not. For this reason our `noisy' initial conditions, which remove the effects of the smallest scales and save computational time are the best choice.

In \autoref{Fig:DiscDens} we show the gas and dust surface density for $\St = 3$ and $\St = 10$, which provides a useful reference for interpretation of the collision velocities. The dust surface density scale is arbitrary due to the use of the test particle limit; for reference we scale the density such that the unperturbed gas and dust surface density are equal. The range $\St =3$ and $\St = 10$ bridges the region where the width of the dust spiral features is the same as those in the gas, with $\St = 3$ showing dense spiral arms and regions where the disc is almost dust-free. As well as dust collecting in individual spirals it is clear that these spirals interact with each other forming kinks in the linear structures.

\section{Measuring the collision velocity}
\label{Sec:DustDisp}

\begin{figure}
 \centering
\includegraphics[width=\columnwidth]{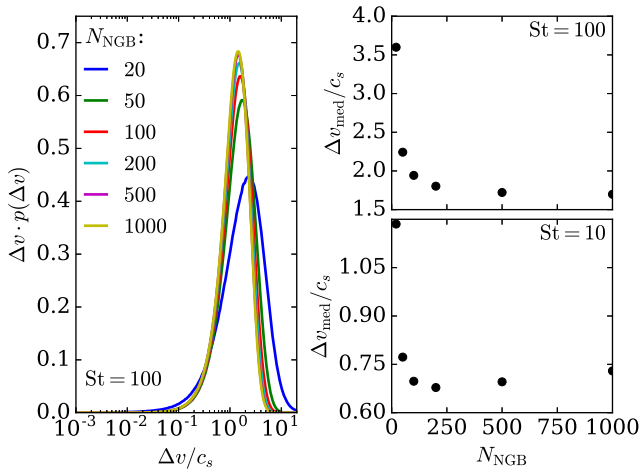}
\caption{\emph{Left}: Velocity distribution of collisions when the number of neighbours used to select the collisions is varied. For a simulation with $10^6$ gas and dust particles and ${\rm St} = 10$. \emph{Right}: Median collision velocity as a function of $N_{\rm NGB}$ for different Stokes numbers. Increasing the sampling volume initially decreases the median collision velocity due to a more accurate gradient subtraction. The slight increase to the large $N_{\rm NGB}$ at $\St=10$ is due to higher order terms in the background velocity.}
\label{Fig:VelPDF_ngb}
\end{figure}

If we denote the probability that two particles separated by a distance $\delr$ have
a relative velocity between $v$ and $v+\delv$ as $P(\delr, v) \delv$, then the
probability density of relative velocities, $p(v)\delv$, is given by
\begin{equation}
p(v)\delv \propto \lim_{\delr \rightarrow r_0} P(\delr, v) g(\delr) \delv,
\end{equation}
where $g(\delr)$ is the probability that two particles are separated by a distance $\delr$ and the constant of proportionality is chosen such the the probability normalises to unity. Here we have neglected the formal spatial dependence of these quantities since we will average over the disc anyway. The limit $r_0$ should be twice the particle size, but since this is much smaller than typical flow scales (mm or cm compared with AU), it is safe to consider $r_0 \approx 0$. Taking the limit $\delr \rightarrow 0$, then  $g(\delr)$ is proportional to the density. For sufficiently small $\delr$, $P(\delr, v)$ and $g(\delr)$ are approximately independent of $r$.  Since $g \propto \rho$, we can measure the relative velocity distribution, $p(v)$, by choosing particles the same way as they are chosen in the density calculation, i.e. using all particle pairs formed from the neighbours within the smoothing volume. To get a good estimate of the distribution one ideally wants to be in the limit of a large number of neighbours and small sampling volume, a regime that is limited by computational expense. It is important to make the distinction that the differential collision rate, $\Gamma(v)$, is different from the relative distribution measured from the simulation, $p(v)$, since the time between collisions $t_c = (n \sigma v)^{-1}$, depends on the velocity. Therefore, the rate of collisions at a given velocity obeys $\Gamma(v) \propto v p(v)$, which we use for calculating the typical collision velocity.

Some improvement can be made over the direct estimate by noting that the relative velocity of two particles separated by a distance $\delr$ includes both a term from the physical velocity dispersion, and a contribution from the background flow. For example, Keplerian shear introduces a velocity gradient into the disc, which produces a velocity difference between two particles of order $\tfrac{\delr}{H} c_s$. As long as  a sufficiently large number of particles is used to measure the velocity gradient, it can be safely subtracted when calculating the collision velocities. This has been verified directly for the case of differing particle sizes, in which we find the same velocity dispersion independently of which particle size is used to remove the background gradient (see \autoref{Sec:BiDisp}). We measure the gradient using the integral approximation for gradients (\autoref{Eqn:IGrad}) and use the corrected relative velocity in measuring the p.d.f.  For the relative velocity we use the 1D equivalent r.m.s. velocity, $\sqrt{\Delta \vec{v} \cdot \Delta \vec{v} / 2}$, rather than the projected velocity, ${\Delta \vec{v} \cdot \Delta \vec{x}} / |\Delta \vec{x}|$, since it appears to be more robust to subtracting off the background gradient. This choice results in $p(\Delta v) \propto \Delta v$ for small $\Delta v$, rather than $p(\Delta v) \approx {\rm const}$. We find the difference in the mean velocities is minor, which are a factor of 2 higher for the r.m.s. case.

To find the optimal sampling volume, we computed the velocity p.d.f. for a variety of number of neighbours, $N_{\rm NGB}$, the results of which are shown in \autoref{Fig:VelPDF_ngb} for $\St = 10$. The behaviour of the measured p.d.f. differs for small ($N_{\rm NGB} \lesssim 100$) and large volumes. For small volumes the p.d.f. is affected by a noisy gradient estimate, which results in a broader distribution as the background velocity is not removed accurately. For large $N_{\rm NGB}$, there are two effects. Firstly, as $N_{\rm NGB}$ increases the contribution to the p.d.f. from particles at the smallest separations is down-weighted. This reduces the width of the low velocity tail in the p.d.f, which is enhanced by particles that are very close to each other and therefore have had very similar histories. For particles with large separations, higher order terms in background velocity become important for $N_{\rm NGB} > 200$ and $\St = 10$, which leads to a change in the mean collision velocity, proportional to $N_{\rm NGB}$ (or $h^2$), of order a few per cent. \autoref{Fig:VelPDF_ngb} shows that the optimal choice will be both a function of resolution and $\St$, but we consider $N_{\rm NGB} = 200$ to be a good compromise.

\section{Sampling scale}
\label{Sec:Sample}

\begin{figure}
 \centering
\includegraphics[width=\columnwidth]{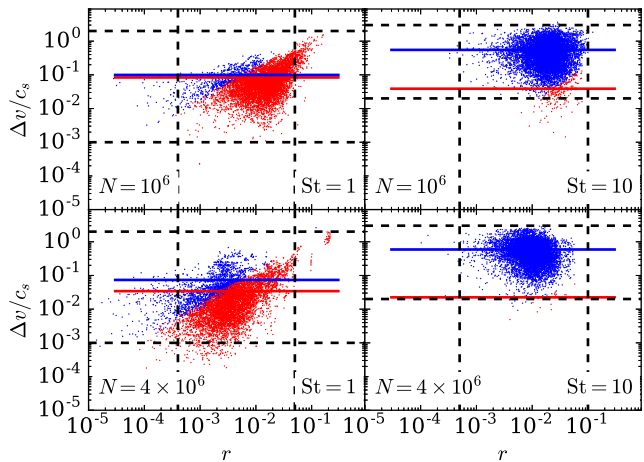}
\caption{Relative velocity for all neighbours of 100 randomly selected particles plotted against the separation between the pair of particles. Results  are shown for two different resolutions and two different Stokes numbers. Blue points refer to particle pairs that are marked as resolved, while red points show unresolved pairs.\
The solid lines denote means for resolved/unresolved pairs and the black dashed lines are included as approximate bounding boxes for the $N = 10^6$ case to guide the eye. Increasing the resolution causes resolved pairs to move to  lower separations at the same velocity; while unresolved pairs move to both lower separations and lower velocities.
}
\label{Fig:V_R_dist}
\end{figure}

In measuring the velocity from simulations we introduce a sampling scale,
$\lambda_{\rm sam}$, into the problem in addition to the stopping distance, 
$\lambda_{\rm stop}$, and driving scale, $\lambda_{\rm d}$. To measure velocity 
dispersion we need
\begin{equation}
\lambda_{\rm sam} < \min(\lambda_{\rm d}, \lambda_{\rm stop}),
\end{equation}
and we have $\lambda_{\rm d} \sim H$. The first inequality can be satisfied
provided the simulation is well enough resolved 
($\lambda_{\rm sam} \ll H$).
$\lambda_{\rm stop}$ is however $\Delta v t_s$ and so there will always be
some particles in the tail of the relative velocity distribution for which 
$\Delta v t_s <$ the particle separation, $r$ ($< \lambda_{\rm sam}$). We 
term these particles `unresolved' because, in the absence of driving by
small scales, the measured relative velocity of 
such particles with respect to a reference particle is an upper limit since in this case
the relative motion will be further damped before the particles 
collide\footnote{In the turbulence literature the resolved/unresolved pairs are
often  referred to as caustic/continuous pairs 
\citep{Falkovich2002,Wilkinson2006,Gustavsson2011,Pan2013}. The name continuous
reflects the continuous variation of the relative velocity as the particles
approach each other, while the name caustic reflects that the orbits cross.
Since in our case the continuous particles are below the resolution limit of 
the simulations, we instead use the term unresolved.}.

For any reference particle, we can sort the $N_{\rm NGB}$ pair-wise interactions within
$\lambda_{\rm sam}$ according to whether $\Delta v t_s < r$ (unresolved) or 
$\Delta v t_s > r$. If $\lambda_{\rm d} \gg \lambda_{\rm sam}$ then we do not 
expect a significant spatial 
gradient of the velocity of resolved particles within $\lambda_{\rm sam}$ (after
having subtracted off the contribution from the mean velocity field). We illustrate
such a case in the right panels of \autoref{Fig:V_R_dist} where we see that by 
construction the resolved particles in this case have a higher velocity 
dispersion than the unresolved. Increasing the resolution clearly only affects
the separation of resolved pairs, while the number of unresolved pairs decreases.
The slight lack of low velocity dispersion pairs at small separations hints at a
density dependence of the velocity distribution, with the higher velocity dispersion
collisions more probable in higher density environments.

When we present our results we will (for each value of
the Stokes number) present particle velocity dispersions considering only the
resolved particles and also those including the unresolved too - evidently
the fraction of unresolved particles varies with both Stokes number
and resolution ($\lambda_{\rm sam}$). The way that such plots 
(\autoref{Fig:MeanVel_St}) should be interpreted is that the resolved values
represent a true representation of the velocity dispersion of that particle
subset; in cases where this value is significantly above the `total' value,
then the latter is an overestimate of the true velocity dispersion because of the
important contribution from unresolved interactions.

In the limit of low $t_s$ we also have to consider the possibility of driving 
of velocity dispersion on small scales ($<\lambda_{\rm sam}$)
due to the relative motion of individual gas particles to which
such dust particles are tightly coupled. In this case we expect  this
purely numerical effect to introduce a gradient in relative velocities
within $\lambda_{\rm sam}$. The upper left panel of \autoref{Fig:V_R_dist}
shows such a case in which there is clear gradient in the collision velocity,
and that resolved particles tend to be close to the target particle and also at
lower velocities. This can result in the magnitude of unresolved velocity
dispersion exceeding that of the resolved particles, which we use in 
\autoref{Sec:MonoDisp} to identify the contribution from noise. 

\section{Mono-disperse case}
\label{Sec:MonoDisp}

\begin{figure}
 \centering
\includegraphics[width=\columnwidth]{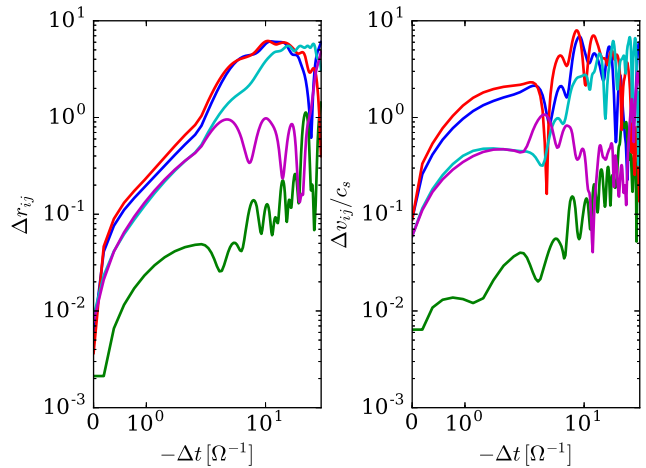} 
\caption{The relative velocity and separation between five randomly selected particles and their nearest neighbours as a function of time into the past, measured in local dynamical times. All particles have $\St = 10$. It is clear that their separation a few dynamical times ago ($-\Delta t \sim t_s$) is a better predictor of the particles relative velocity than their current separations, which are all similar.}
\label{Fig:Vel_sep}
\end{figure}

\begin{figure*}
 \centering
\begin{tabular}{cc}
\includegraphics[width=\columnwidth]{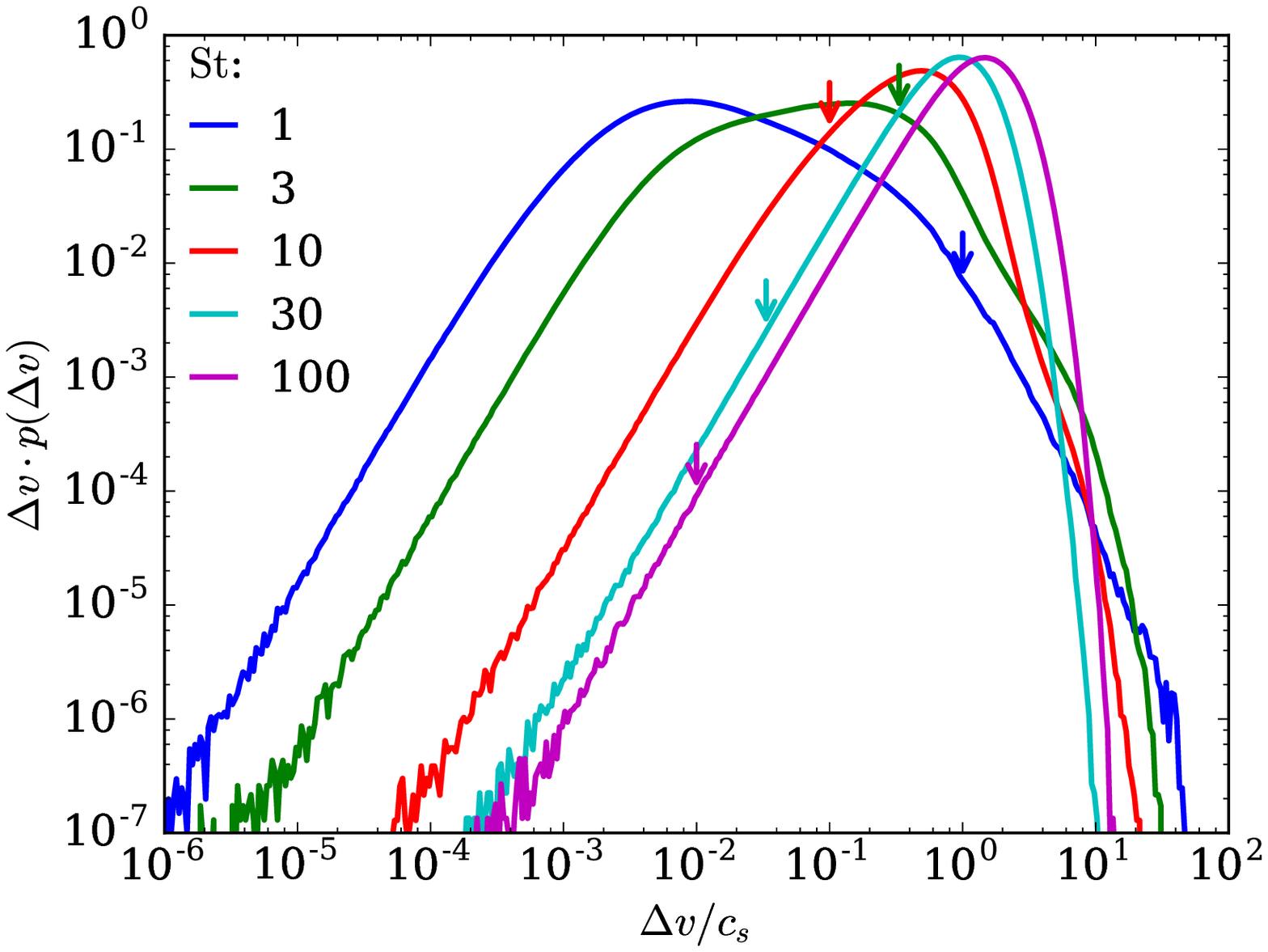} &
\includegraphics[width=\columnwidth]{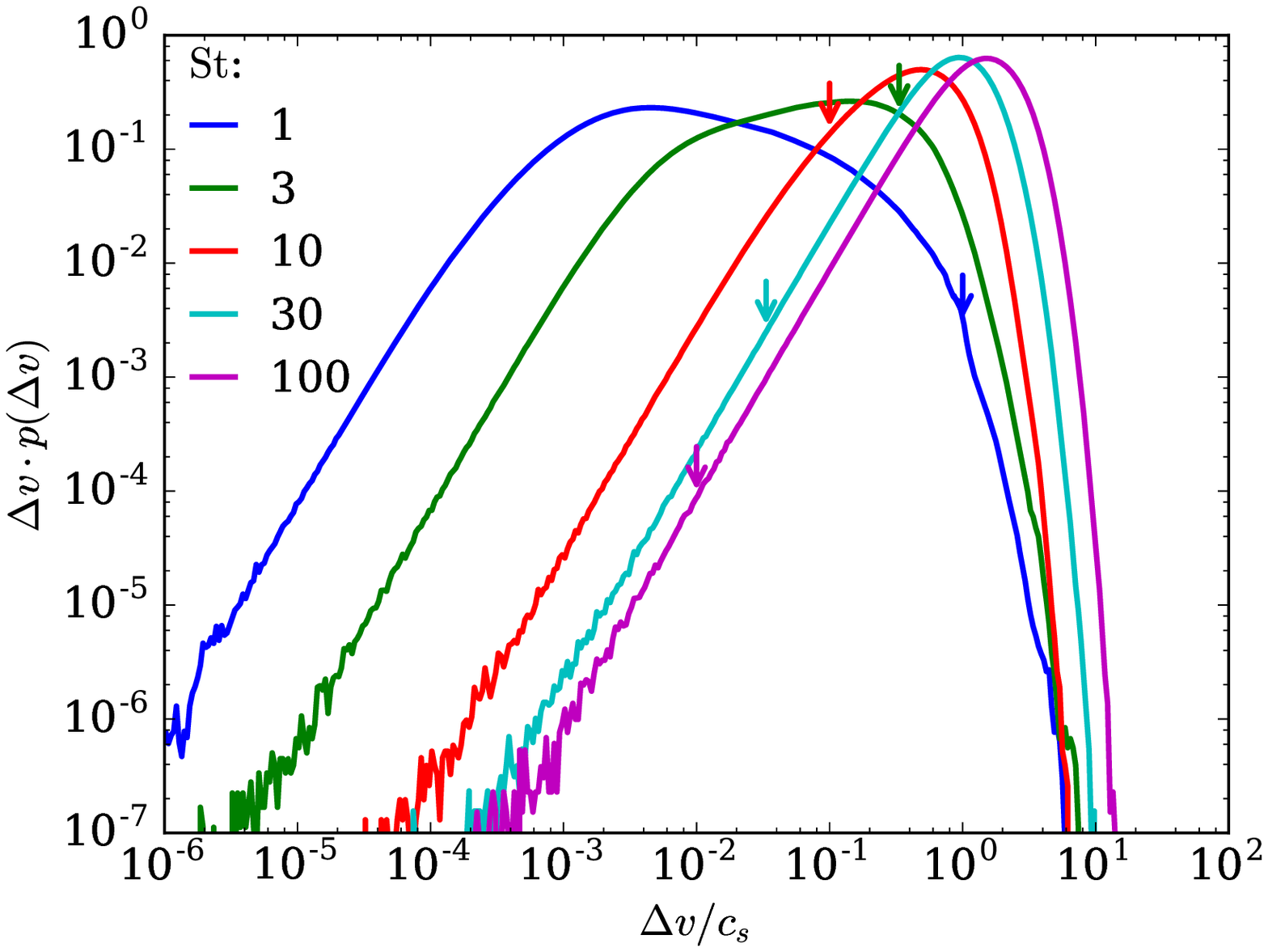}
\end{tabular}
\caption{\emph{Left}: Collision velocity distribution for all particles at a range of different Stokes 
numbers. \emph{Right}: The same, but with dust particles in low density environments excluded, i.e only those particles with $\Sigma_{\rm D} > \Sigma_{\rm G}$ are used (where $\Sigma_{\rm D} / \Sigma_{\rm G} = 1$ is the global average). The results are shown for simulations run at a resolution of $4 \times 10^6$ particles per phase. The coloured arrows show the velocity for which $\lambda_{\rm stop} = H$.
}
\label{Fig:PDF_St}
\end{figure*}

We first examine in detail the idealized case of equal-sized particles, thereby
neglecting effects that arise due to differences in stopping time, such as 
radial and azimuthal drift. Since in a smooth disc equal-sized particles have 
zero relative velocity the mono-disperse case is ideal for determining the 
random component of the collision velocity. In \autoref{Sec:BiDisp} we discuss
the case of collisions between different particle sizes. 

In \autoref{Fig:Vel_sep} we show the separation and relative velocity between pairs of $\St = 10$ particles as a function of time into the past. For each of 5 randomly selected particles we show the history of the separation of the particle's current nearest neighbour. It is clear that while all particle pairs have a similar separations at the current time, they have quite different relative velocities. The relative velocity is much more accurately predicted by the separation of the particles $3$--$10\Omega^{-1}$ ago, i.e. within of order $t_s$. This neatly demonstrates our argument above that the spatial separation of particles is important for determining the relative velocity of particle pairs, and also that information about the particles separation far into the past has been lost. We will thus discuss the measured collision velocities in terms of this model in the following sections.

\subsection{Dependence on Stokes Number}
\label{Sec:StDep}

\autoref{Fig:PDF_St} shows the distribution of relative velocities for $\St \ge 1$; this plot contains only resolved particles (unresolved particles are a minority component at these Stokes numbers). At low velocity, the distribution shows a $ p(v) \propto v$ behaviour, which is a consequence of using the r.m.s. relative velocity. (The lowest Stokes number runs, $\St = 0.1$ and $\St = 0.3$, are not shown as the velocity distribution becomes dominated by unresolved pairs.) Even at large $\St$ we find that velocity distribution is not well fit by a Gaussian and that for $\St = 100$ we find a reasonable match to $p(v) \sim v \exp( - C |v|^{1.3})$. Surprisingly, this is close to the distribution of dust collision velocities in turbulent gas, $p(v) \sim \exp( - C |v|^{4/3})$ \citep{Gustavsson2008}, despite the additional driving by gravitational forces.

The velocity distribution for $\St \le 3$ shows a power-law tail to large velocities that is due to collisions involving particles at low densities in the simulation. Even though there are relatively few particles in low density regions (especially for small $\St$), they can overlap dense regions containing many particles and thus make a significant contribution to the relative velocity distribution. Since these particles can have separations larger than a scale height the relative velocity may change significantly before a collision occurs. Furthermore, the gradient subtraction process actually increases the strength of these tails, most likely because the large number of closely associated particles in the high density region biases the gradient estimate. We therefore suggest that they are unphysical, and have found that if we exclude collisions in which either of the particles are in low density environments ($\Sigma_{\rm D}/\Sigma_{\rm G}$ less than the global average, which we define as $\Sigma_{\rm D} / \Sigma_{\rm G} = 1$ by scaling the dust density) then these tails are reduced (right hand panel, \autoref{Fig:PDF_St}) and the distributions show a similar cut off to the high $\St$ distributions. The excluded fraction of particles is greatest for the high $\St$ cases (less than 10 per cent), but we find even in this case it makes little difference to distribution apart from the high velocity tails. However, particular choices of the cut-off density, make very little difference to the results because the dust particles are preferentially found in features with ${\Sigma_{\rm D}/\Sigma_{\rm G} > 1}$.

\begin{figure}
\centering
\includegraphics[width=\columnwidth]{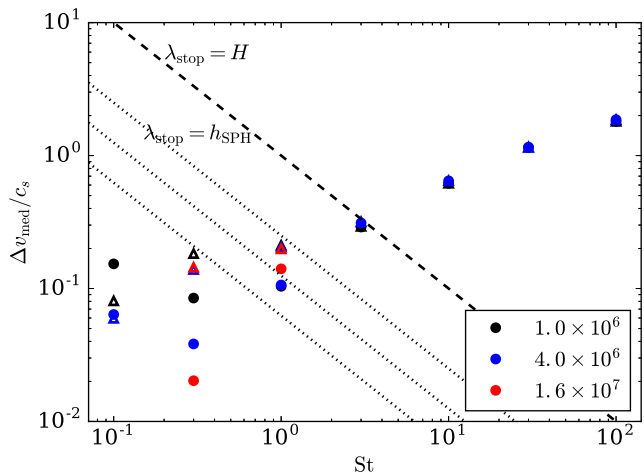}
\caption{Median collision velocity as a function of Stokes number for both the full distribution (filled circles) and resolved particles only (open triangles). Particles in low density regions have been excluded. The velocity for which the stopping distance $\lambda_{\rm stop}$ equals the driving scale, $H$ and the resolution of the simulations $h_{\rm SPH}$ are shown by the dashed and dotted lines, respectively. (The $h_{\rm SPH}$ shown is for the gas, $h_{\rm SPH}$ for the dust is $\St$ dependent, but generally within a factor of 2 of that of the gas.) Note that for $\St = 0.1$ the median velocity of the resolved particles falls below the median velocity of all particles, indicating driving by numerical noise.
}
\label{Fig:MeanVel_St}
\end{figure}

\begin{figure}
\centering
\includegraphics[width=\columnwidth]{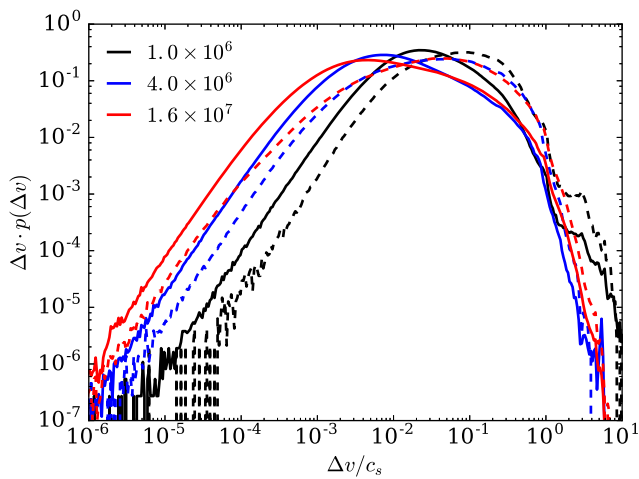}
\caption{Collision velocity distribution for particles with $\St = 1$ at different resolutions. The solid
lines refer to the total distribution, while the dashed lines  show the distributions of resolved
collisions only. Particles in low density environments have been excluded.}
\label{Fig:PDF_Caustic}
\end{figure}

\begin{figure*}
\centering
\includegraphics[width=\textwidth]{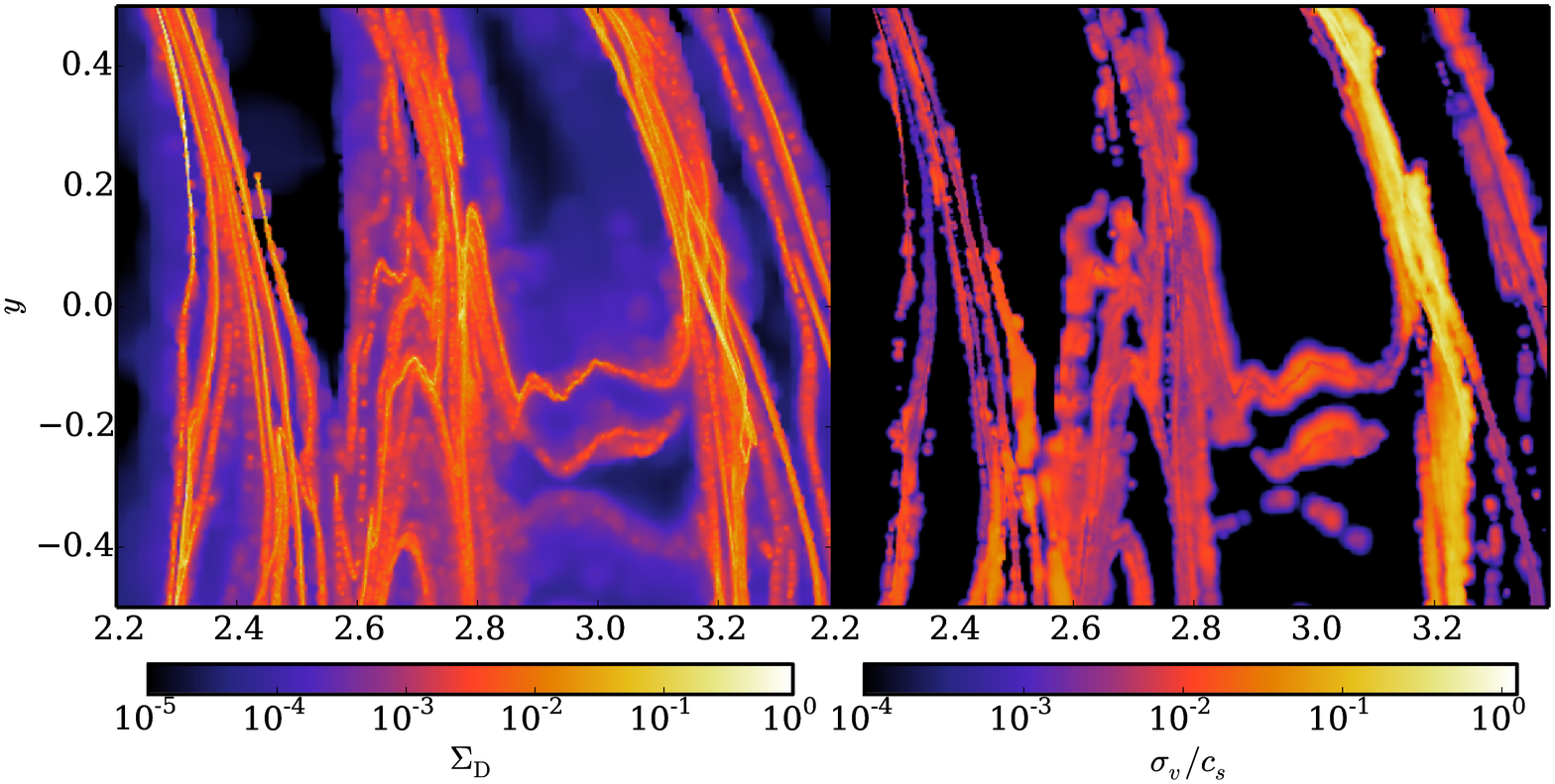} \\
\includegraphics[width=\textwidth]{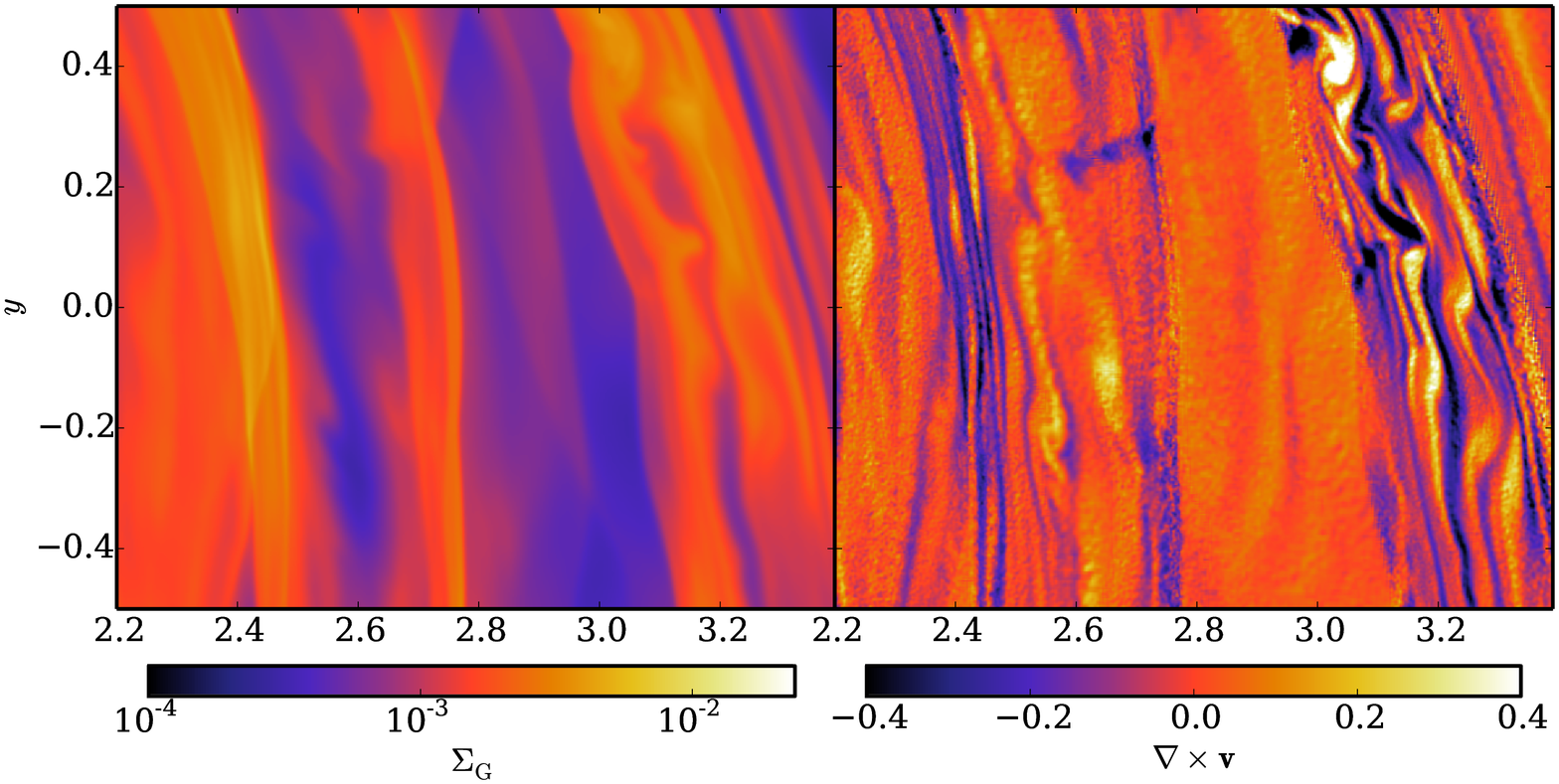}
\caption{\emph{Top}: Spatial distribution of the dust density (left) and velocity dispersion (right) for $\St = 1$ in a small region of the disc. The highest velocity dispersion environments are associated with regions where the narrow spiral features in the density cross. In those regions where the spirals merely closely approach the velocity dispersion can be considerably smaller. The particles flagged as in low density regions have been excluded from the velocity dispersion calculation, resulting in $\sigma_v = 0$ in the lowest density regions. \emph{Bottom}: Gas surface density and vorticity (with the keplerian background subtracted) in the same region. The spiral crossing region in the dust is associated with high vorticity in the gas, suggesting that the largest velocity dispersion at $\St=1$ is driven by turbulent velocities rather than gravitational perturbations.}
\label{Fig:SigV_map}
\end{figure*}

\begin{figure}
\centering
\includegraphics[width=\columnwidth]{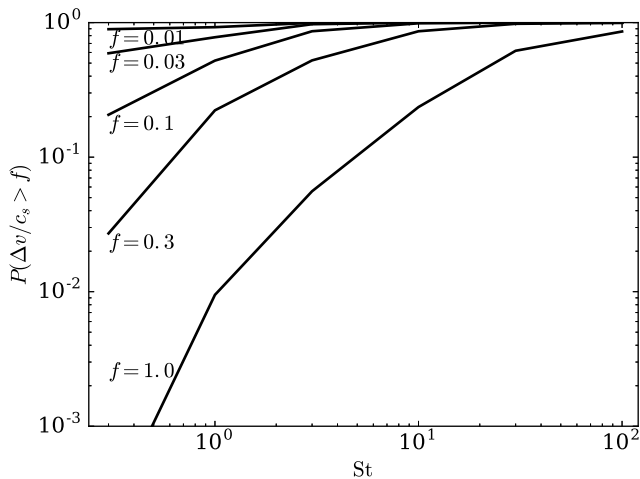} 
\caption{Fraction of collisions that occur above a given velocity threshold for different Stokes numbers.}
\label{Fig:Vel_frac}
\end{figure}

\begin{figure*}
\centering
\begin{tabular}{cc}
\includegraphics[width=\columnwidth]{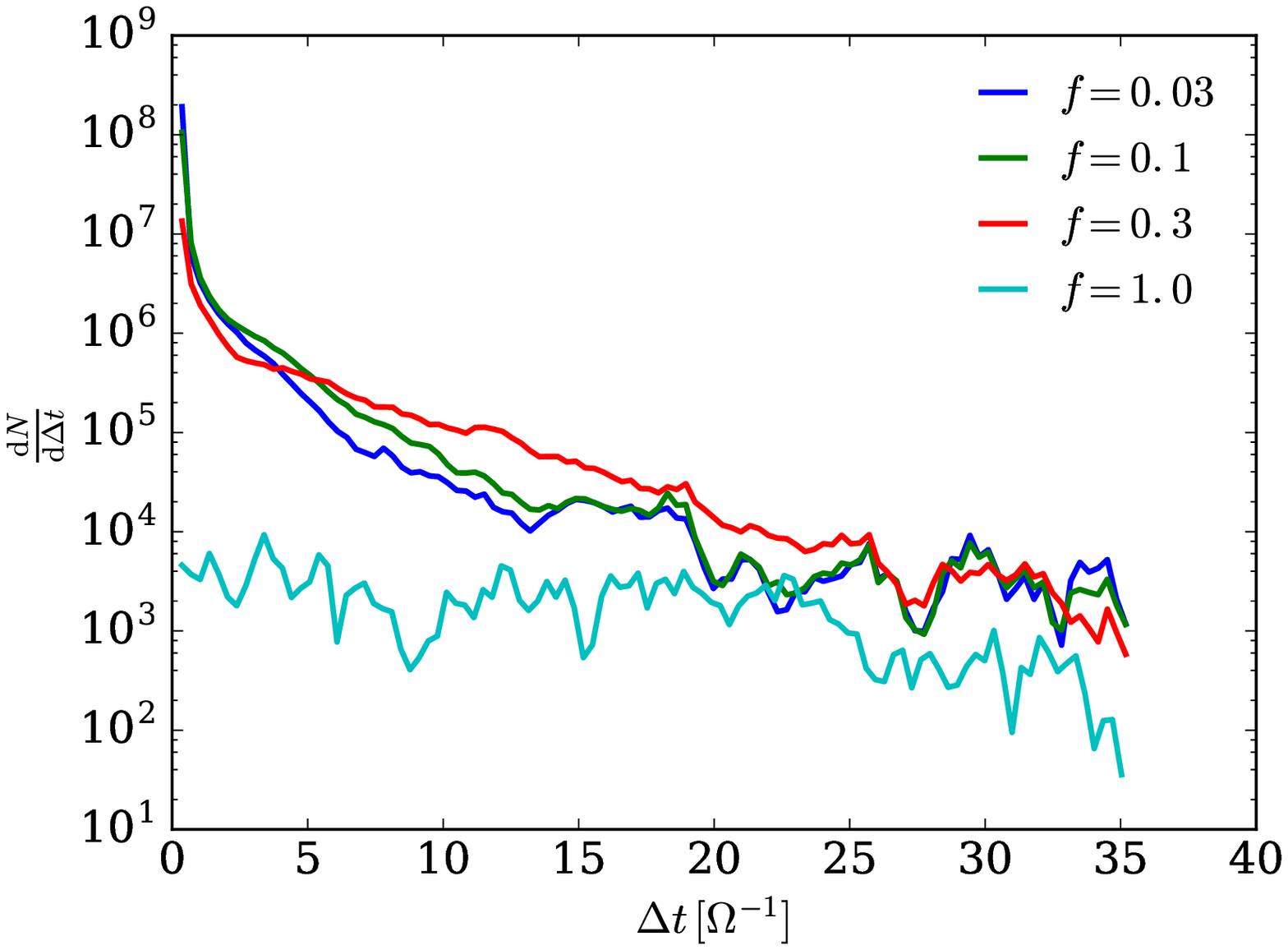} &
\includegraphics[width=\columnwidth]{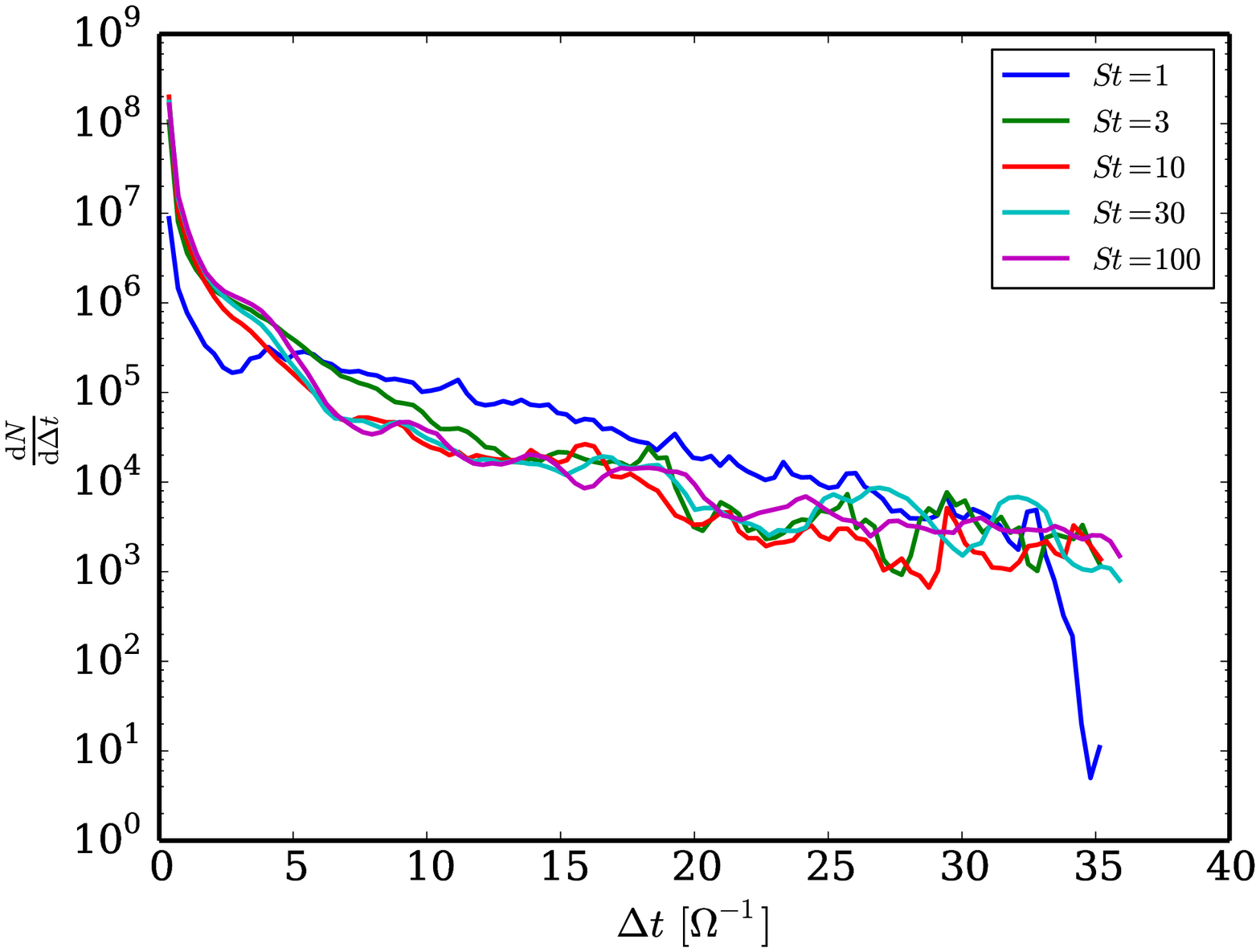}
\end{tabular}
\caption{Distribution of times between particles being in environments with typical collision velocities above a given fraction of the sound speed. \emph{Left}: Distribution for $\St = 3$ and varying threshold velocities, $f = \Delta v / c_s$. \emph{Right}: Distribution for $f = 0.1$ and varying \St.}
\label{Fig:Vel_frac_dist}
\end{figure*}

\autoref{Fig:PDF_St} also indicates for each Stokes number the velocity scale at which $\lambda_{\rm stop}= H$ and we see that for $\St \ge 10$ the majority of collisions occur in the regime where $\lambda_{\rm stop} > H$. This is consistent with velocity driving on a scale $H$, as is expected for gravitational driving (see the gas surface density snapshot in the lower left of \autoref{Fig:SigV_map} which indicates the broad striations on scale $H$). 

The median collision velocity, $\vmed$, is shown in \autoref{Fig:MeanVel_St}, and is showing shows signs of saturating towards 2 to 3$c_s$ by $\St = 100$. In the absence of drag, the sound speed affects the velocity distribution only in as far as it controls gravitational perturbations from the disc: these are jointly
set by $\beta$ \citep{Cossins2009} and by the disc surface density, $\Sigma$, which is related to $c_s$ via the constancy of the Toomre Q parameter. For $\St \ge 3$ we see that \vmed{} is converged even at the lowest resolution, which is expected since $\lambda_{\rm stop} \gg h$. Between $\St = 3$ and $\St = 100$ we find $\vmed \propto \St^{1/2}$, the same dependence that is found for particles in turbulence in the inertial range \citep[e.g.][]{Ormel2007}. This behaviour may indicate that the particles are undergoing a random walk in velocity space, which would produce a $\vmed^2 \propto \dot{Q} t_s$ behaviour for a heating rate, $\dot{Q}$, that is independent of $\vmed$ and $t_s$. As discussed above, this is as expected for gravitational driving of particles in this range of St for which $\lambda_{\rm stop} > H$.

Turning now to the behaviour at lower $\St$, we see from Figures~\ref{Fig:MeanVel_St} and \ref{Fig:PDF_Caustic} that $\Delta v_{med}$ is only approximately converged at $\St = 1$  by $10^7$ particles per phase, and for $\St = 0.1$ a converged velocity distribution would require $> 10^9$ particles. This is because $\lambda_{\rm stop}$ is a very strong function of $\St$ since both  $\vmed$ and $t_s$ depend on $\St$. We also see from Figures~\ref{Fig:PDF_St} and \ref{Fig:PDF_Caustic} that the velocity distribution broadens (relative to its form at high $\St$) at a Stokes number of around 3 and that a significant fraction of particles have velocities for which $\lambda_{\rm stop} < H$, which is an unexpected outcome in the case of gravitational driving. This behaviour is explained by examination of the lower right panel of \autoref{Fig:SigV_map}, which shows regions of localised structure in the gas vorticity field on scales $< H$. These regions are associated with locations where spiral arms cross (as can be seen in the distribution of $\St=1$ particles in the upper left panel) and these drive locally high values of the velocity dispersion of such particles (upper right panel of \autoref{Fig:SigV_map}). We emphasise a qualitative shift in particle driving at around $St = 3$, where $\lambda_{\rm stop} = H$. At higher Stokes number, all particles are subject to gravitational driving and the velocity field reflects a balance between such driving and gas drag. At lower Stokes number, the dominant {\it driving} is also by drag forces whose strength is determined by the local gas velocity field. The majority of particles which are trapped in spiral features experience a quiescent velocity field and have very low mutual velocities which we struggle to resolve. A small fraction of particles however experience strong driving in regions of high vorticity where spiral features cross. We are able to measure a converged median collision velocity for particles in the latter category at $St=1$ and are approaching convergence for this particle subset at $St=0.3$ (triangles in \autoref{Fig:MeanVel_St}).

At $\St = 0.1$ we find that the relative velocity of resolved particles is smaller than that of the unresolved ones. The explanation for this is that the resolved particles are on average closer than the unresolved ones and the reason that these particles are flagged as `resolved' is thus due to their small separation. This results in a lower than average velocities since the particles are sufficiently close that they share the same gas neighbours and see the same realisations of the noise. This inversion of the resolved and unresolved velocity dispersions is a clear signature of noise induced by jitter of individual gas particles. This effect becomes less important at higher Stokes number because the noise is averaged over more gas particles and more time-steps. We are confident that this interpretation is correct since the same behaviour can be clearly seen in test problems (see Fig.~12 of \citealt{Booth2015}), where the velocity dispersion can be unambiguously determined to be due to noise. By extrapolating \vmed{} from  $\St = 0.1$ and $\St = 0.3$ it is clear that noise is not contributing to $\vmed$ at $\St = 1$.

Since for $\St < 3$ we find the presence of both  high and low velocity collisions, we also consider the fraction of collisions that occur above a given threshold, which is important for examining the growth and fragmentation of dust grains. The results are shown in \autoref{Fig:Vel_frac}. At $\St = 1$ more than 30 per cent of collisions occur at velocities greater than $0.1c_s$ and it is unlikely that this drops below 10 per cent until $\St \lesssim 0.1$. However, at $\St = 1$ the fraction of collisions with velocities greater than $c_s$ ($\sim 10 \vmed$) is less than 1 per cent, due to the exponential tail in the velocity distribution.

Even though a small fraction of high velocity collisions occur, if particles can avoid environments in which large collision velocities occur then it may also be possible for particles to avoid fragmentation for long periods of time. We have investigated this by examining the distribution of delay times between particles entering high velocity dispersion environments. This was done by identifying the local velocity dispersion of each particle in simulation snapshots separated by an inner dynamical time scale ($\Omega(R_{\rm in})^{-1}$), once a $\Delta t$ had been found it was scaled to the local dynamical time-scale. We limit the investigation to particles within the range $r = [2, 3]$, where the effective time resolution of the snapshots is 0.2 to $0.35\Omega^{-1}$. The distributions are shown in \autoref{Fig:Vel_frac_dist}. We find that at large $\Delta t$, the distribution is in good agreement with an exponential decay. For $\Delta t \lesssim 3\Omega^{-1}$ the distribution is clearly enhanced relative to the tail. For the range of $\St$ considered here, the typical delay times are considerably shorter than the collision time ($t_c > 10^4 \Omega^{-1}$ at $\St = 3$, \autoref{Eqn:CollTime}). Although particle trapping in spiral features may significantly reduce the collision time (for $\St \sim 1$), it is likely that the a given dust particle samples the disc-averaged collision velocity distributions fairly evenly.

\begin{figure}
\centering
\includegraphics[width=\columnwidth]{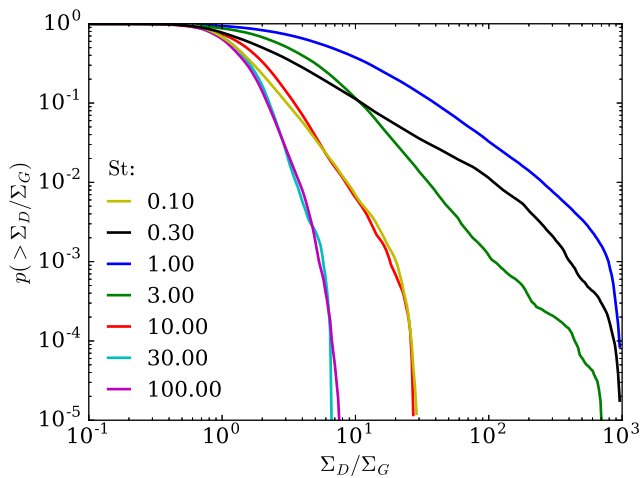}
\caption{Enhancement of dust density relative the gas density. The $y$-axis 
shows the probability that a given dust particle is found in a region that
has an enhancement above a given factor of the background value.}
\label{Fig:PDF_Dens}
\end{figure}

Finally, although the focus of our analysis has been the quantification of the velocity dispersion in relation to solid growth by collisions, we also briefly consider the issue of planetesimal assembly by direct gravitational collapse of the dust. For dust self-gravity to become strong enough that dust driven collapse to form planetesimals to occur the dust must be concentrated in regions where the local dust density is comparable, or higher than the local gas density. We find that the factor $\sim 100$ enhancement required for this to happen is limited to the range $0.3 \lesssim \St \lesssim 3$ (\autoref{Fig:PDF_Dens}). The effects of dust self-gravity, which have been neglected here may increase this range slightly, but equally once the dust mass becomes locally comparable to the gas mass the drag force will weaken (due to a factor of $(1 + \rho_{\rm D} / \rho_{\rm G})^{-1}$ in the stopping time), which may also reduce the viable range of $\St$ for $\St < 1$.

\subsection{Dependence on $\beta$}
\label{Sec:BetaDep}

\begin{figure}
\centering
\includegraphics[width=\columnwidth]{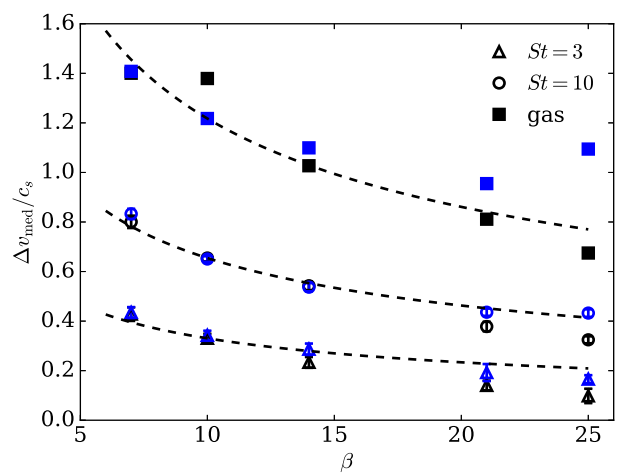} 
\caption{Median collision velocity for dust as a function of cooling time, $\beta$, at resolutions of $10^6$ (black) and $4\times 10^{6}$ (blue) particles per phase. The r.m.s gas velocity, $\langle (v_g/ c_s)^2 \rangle^{1/2}$, and dashed curves that denote $\beta^{-0.5}$ are also shown. The error bars denote the 1-$\sigma$ snapshot to snapshot variation}
\label{Fig:MeanVel_Beta}
\end{figure}

\begin{figure}
\centering
\includegraphics[width=\columnwidth]{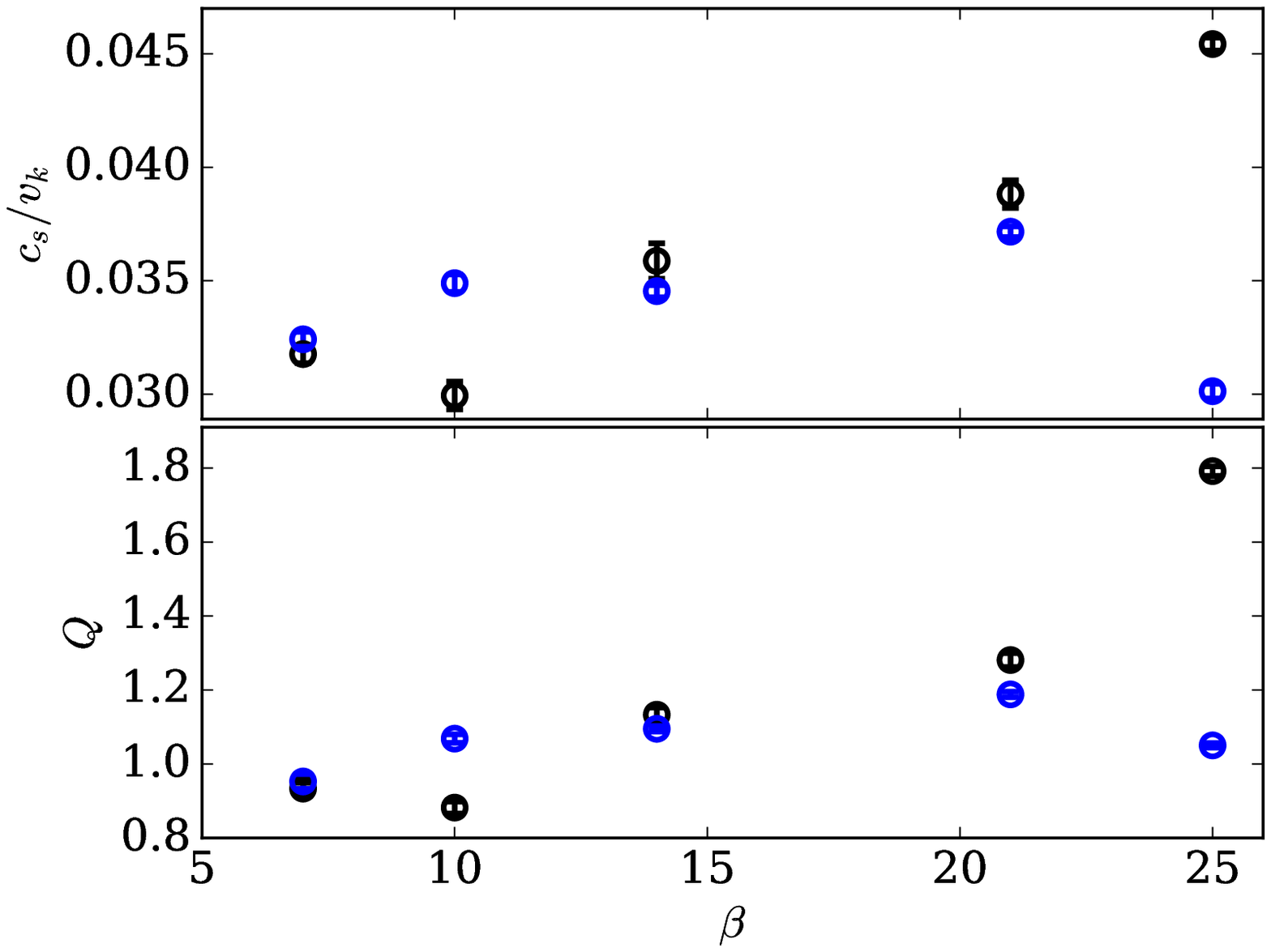} 
\caption{Ratio of the mean sound speed to the local Keplerian velocity and Toomre $Q$ parameter at resolutions of $10^6$ (black) and $4\times 10^{6}$ (blue).The error bars denote the 1-$\sigma$ snapshot to snapshot variation.}
\label{Fig:cs_Q}
\end{figure}

Our canonical value of the cooling time $\beta = 10$ represents cooling just inside the boundary for gas phase fragmentation, where the perturbations are strong. Due to the strong dependence of $\beta$ on radius \citep[$\beta \propto r^{-4.5}$,][]{Rafikov2005, Clarke2009}, the surface density perturbations will be weaker at smaller radii, which will affect the collision velocities. In order to investigate this we have run simulations  in which we varied the cooling time, $(\beta \Omega)^{-1}$ (at resolutions of $10^6$ and $4\times 10^6$ particles per phase). We have measured the velocity dispersion for particles with $\St = 10$ and $\St = 3$, for which the driving scale should be well resolved.

The results for the velocity dispersion averaged over 50 inner dynamical times are shown in \autoref{Fig:MeanVel_Beta}, along with the r.m.s. gas velocity. We also show curves of $\beta^{-1/2}$, the dependency expected for the r.m.s. gas velocity perturbation when heating by weak shocks ($\propto v^2$) balances cooling ($\propto \beta^{-1}$). This dependence was predicted by \citet{Clarke2009} in order to explain the $\Delta \Sigma / \Sigma \propto \beta^{-0.5}$ dependence found in the simulations of \citet{Cossins2009}. We find a slightly steeper dependence of the r.m.s. gas velocity at low resolution, but a shallower dependence at high resolution. Figure \ref{Fig:cs_Q} shows that differences may partly be accounted for by variations in $c_s$, which increases with $\beta$, with the Toomre $Q$ parameter showing similar dependence. We note that a slight increase in $Q$ with $\beta$ should be expected since $Q$ controls the strength of the self-gravity and thus by regulating $Q$ the disc is able to maintain an equilibrium temperature for different $\beta$. However, the behaviour for $\beta \gtrsim 20$ is likely to be affected by artificial viscosity since in 3D shearing box simulations \citet{Shi2014} found only a very weak dependence of $c_s$ on $\beta$ for $\beta > 10$.

Turning now to the median collision velocity for dust grains, we also see a behaviour that is broadly similar to, although slightly steeper than, the $\beta^{-0.5}$ behaviour exepcted for gas. Interpretting this in terms of a random walk that produces $\vmed \propto \St^{1/2}$, we see that the rate of excitation $\dot{Q} \sim \beta^{-1} \propto (\Delta \Sigma / \Sigma)^2$. The fact that we see similar behaviour for the dust and gas is indicative of a single mechanism being responsible for driving both of them -- for the gas this mechanism is spiral density waves that contain an energy density contained $\propto (\Delta \Sigma / \Sigma)^2$ \citet{Cossins2009}. Thus we suggest that it is the energy available in these spiral waves that is responsible for exciting the dust collision velocities, at least for the case $\lambda_{\rm stop} > H$.

\subsection{Density dependent Epstein drag law}

We now revisit the collision velocity with a prescription that takes into account the explicit density dependence of the Epstein drag law instead of fixed Stokes number (i.e. we allow the Stokes number to vary according to the inverse ratio of the local surface density to its azimuthally averaged value). As described in \autoref{Sec:Method}, under the approximation of vertical hydrostatic equilibrium the Stokes number depends only on the surface density. To ensure self-similarity we have normalized the surface density to background (unperturbed) density, to keep the same Stokes number at all locations while including the density dependence.

\begin{figure}
\centering
\includegraphics[width=\columnwidth]{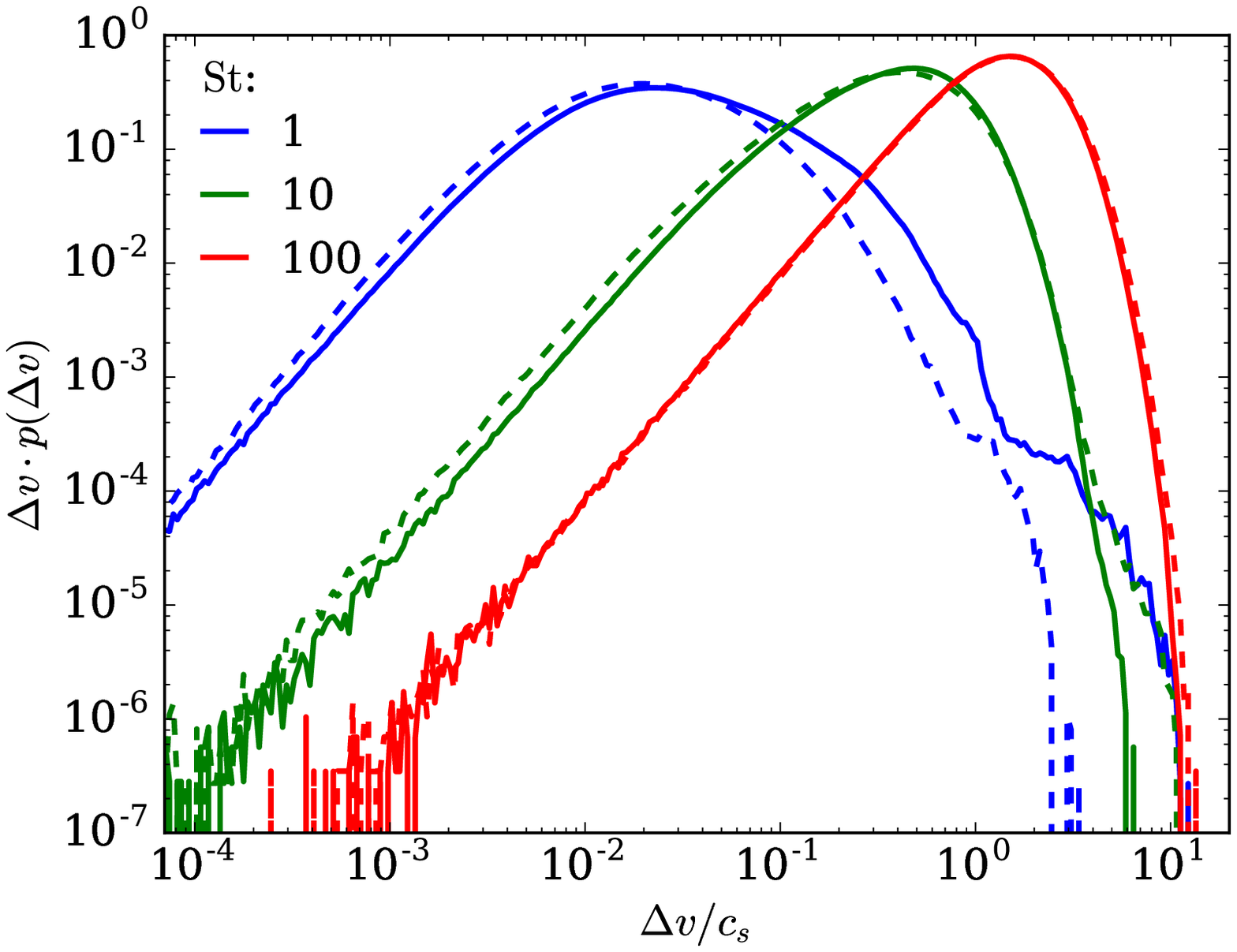}
\caption{Collision velocity distribution for simulations using drag laws based upon fixed Stokes number
(solid) and a self-similar Epstein drag law (dashed).}
\label{Fig:Velpdf_Sigma}
\end{figure}

The effect that the drag law makes on the relative velocity can be seen in \autoref{Fig:Velpdf_Sigma}. For all Stokes numbers $\St > 1$, the differences in the distributions are small and decreasing as the Stokes number increases, with the difference in the mean velocity less than 20 per cent at $\St = 1$. In all cases the mean collision velocity is slightly smaller when using the Epstein drag law, but the difference is small enough that the drag law is unlikely to have a large effect on the outcome of collisions derived from simulations. We interpret this result as being due to the fact that stronger perturbations are associated with higher density regions and the stopping time in these regions will be smaller for the Epstein drag case than the average Stokes number case.

\section{Bi-disperse case}
\label{Sec:BiDisp}
\begin{figure*}
\centering
\begin{tabular}{cc}
\includegraphics[width=\columnwidth]{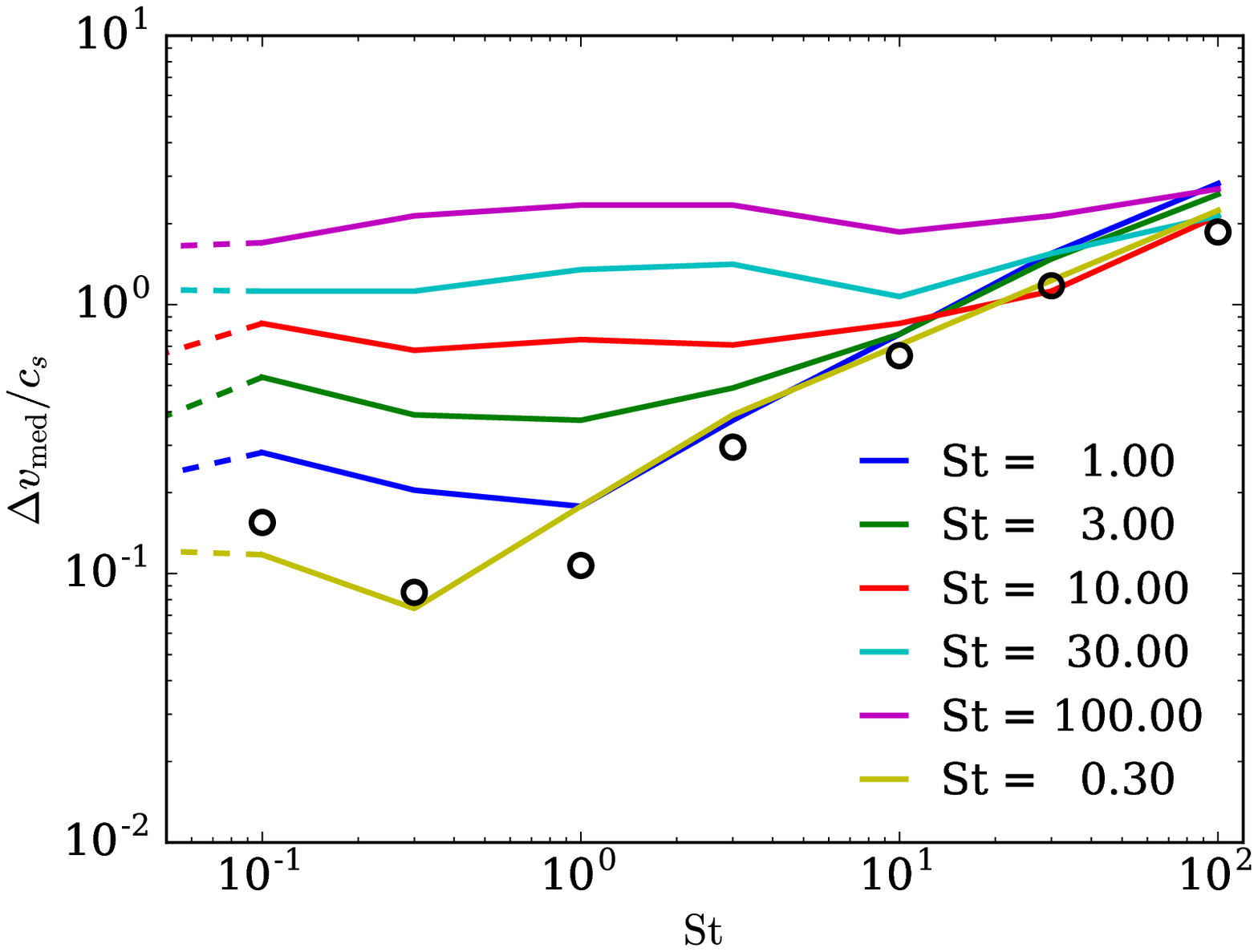} &
\includegraphics[width=\columnwidth]{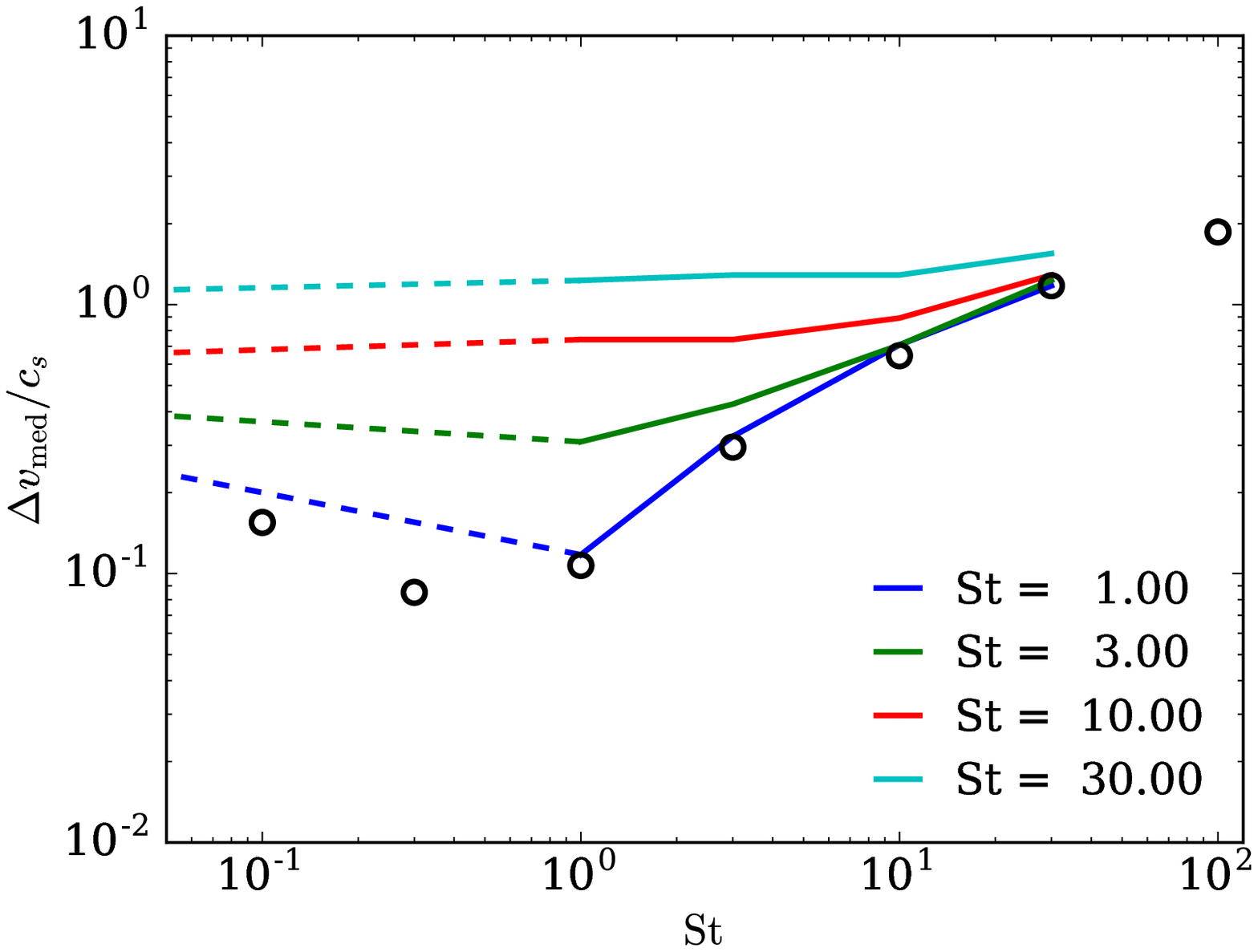}
\end{tabular}
\caption{Median collision velocity for particles in the bi-disperse case. \emph{Left}: Dust resolution 
$2.5 \times 10^5$ particles and gas resolution $10^6$ particles. \emph{Right}: Dust and gas resolution
$10^6$ particles, but for fewer particle sizes. The hollow circles show the results for the mono-disperse
case at a resolution of $10^6$ particles per phase. The curves show the collision velocity between particles
of fixed Stokes number as a function of the Stokes number of the colliding particle. The dashed lines connect
the bi-disperse case to the relative velocity between the dust and gas (a proxy for dust particles in the limit $\St \rightarrow 0$), plotted at $\St = 0.05$.}
\label{Fig:Vel_bidisp}
\end{figure*}

We now turn our attention to the case of collisions between dust particles of different sizes. To ensure that
we are comparing particles in identical environments we have run a pair of simulations using a gas resolution of $10^6$ particles and dust resolutions of $2.5 \times 10^5$ particles per phase and $10^6$ particles per phase at Stokes number $\St = \{0.1, 0.3,1,3,10,30,100\}$ and $\St = \{1,3,10,30\}$  respectively. The collision velocity is then determined for each size by finding the neighbouring  particles of a given size and subtracting off the gradient as before. This process is not symmetric, as only the gradient in one particle size is taken into account, but we find a less than 10 per cent difference on swapping the particle sizes, which confirms that the subtraction effectively removes the background velocity field.

The results of these simulations are shown in \autoref{Fig:Vel_bidisp}. As before, the behaviour
can be broadly split into two categories. For particles with 
$\lambda_{\rm stop}= \vmed t_s \gtrsim H$  ($\St \gtrsim 3$)
the results are consistent with the relative velocity of these particles being determined by the
difference between two random velocity distributions, as expected for particles that undergo many
gravitational perturbations within a stopping time. This result holds even when one of the species
has $\lambda_{\rm stop} \ll H$ since the collision velocity is dominated by the collision velocity of the
larger species. These results reflect those found for particles with $\St > 1$ in turbulent
media \citep{Volk1980,Ormel2007,Pan2010}.

When both particles have $\lambda_{\rm stop} \lesssim H$, the behaviour is different. For nearly identical particles, the velocity is dominated by the (approximately) mono-disperse velocity dispersion and rapidly decreases in strength for $\lambda_{\rm stop} < H$. However, for particles with different Stokes number the relative velocity increases with increasing size difference. This behaviour is driven by the different terminal velocities between particles of differing sizes $\Delta v_{\rm term} = (t_{s,1} - t_{s,2}) \nabla P / \rho$. In this case the \emph{local} velocity dispersion can be much smaller than the global distribution that arises from the differing pressure gradients in the disc.

It is interesting to compare the relative velocity associated with this drift to the radial drift due to the background (azimuthally averaged) disc structure. For a power-law surface density distribution at $Q \approx 1$, the radial drift velocity of a particle is given by 
\begin{equation}
v_r(\St) = - 2 \eta v_K \frac{\St}{1+\St^2} = 3 (1+k) c_s \frac{H}{R} \frac{\St}{1+\St^2}
\end{equation}
where $\eta = -\frac{1}{2 \Sigma r \Omega^2}\tderiv{P}{r} = -3/2 (1 + k) (H/R)^2$,
and $k=-2$ is the power law index of the surface density and we have assumed a razor thin disc to represent the simulations \citep{Whipple1972,Weidenschilling1977,Nakagawa1986}. For our chosen disc parameters, $H/R \approx 0.03$ at $Q = 1$ and $v_r \approx 0.09 c_s \St/ (1 + \St^2)$. The corresponding difference between the unperturbed \emph{gas} rotation velocity and the local Keplerian value, $\eta v_K \approx 1.4 \times 10^{-3} v_K$ ($\approx 0.05 c_s$) is approximately $7\unit{m\,s}^{-1}$ at $30\unit{au}$ (cf. $\sim 50\unit{m\,s}^{-1}$ throughout the minimum mass solar nebular, \citealt{Johansen2014}). From the simulations we find the dust-gas relative velocity is $3.5\times 10^{-2}$, $7.5\times 10^{-2}$ and $17 \times 10^{-2} c_s$ for $\St = 0.1$, 0.3 and 1, suggesting that density perturbations result in drift velocities a few times larger than the background disc.

Since the surface density perturbations, $\Delta \Sigma/ \Sigma \sim 1/\sqrt{\beta}$, we can expect the radial
drift of the background disc to dominate the spiral perturbations for $\beta \gtrsim 100$, thus limiting
the region in which the self-gravitating structure affects small grains. Once the radial drift of the background disc dominates the relative drift in the spirals we can expect the self-gravitating spiral structure to be no longer able to trap dust particles, even at $\St = 1$. This appears consistent with the results of \citet{Gibbons2012}, in so far as they only ran one simulation at the very edge of the region of $v_r$\nobreakdash--$\beta$ space in which radial drift should be important. That simulation showed a mild suppression of the concentration of dust in spiral features by radial drift, supporting our argument.

\section{Discussion}
\label{Sec:Discuss}

We have run smoothed particle hydrodynamics simulations of dust particles in self-gravitating protoplanetary discs, with the aim of understanding whether the conditions for direct gravitational collapse can occur in dust trapped in the spiral arms of the disc, or whether collisions between dust grains lead to fragmentation, thus preventing growth to large enough sizes. To this end we have measured the collision velocity between pairs of particles for a range of sizes, along with the density enhancement. 

We find that while gas self-gravity can drive velocity dispersions of order the sound speed in weakly coupled particles ($\St > 10$), for $\St \lesssim 3$ the largest scales that can effectively drive the velocity dispersion, $\lambda_{\rm d}$, drop below the disc scale height, $H$. Once this happens, driving by the disc gravity becomes inefficient, and the dominant driving mechanism comes from the gas drag. We find that at this transition the fraction of collisions occurring at velocities exceeding a significant fraction of the sound speed drops rapidly. In addition we see that the highest velocity collisions become associated with the crossing of spiral features in the dust density, while collisions between particles within the same spirals occur at considerably lower velocity.

 We find that particles are effectively trapped in spiral features, leading to high local dust to gas ratios, in the range of Stokes number $\sim 0.3$--3. These results are in good agreement with those of \citet{Gibbons2012}, who also found that particles with $\St = 0.1$ or $\St = 10$ are not effectively trapped by the spiral features\footnote{Trapping is most effective when the radial drift velocity is large, i.e. close to $\St = 1$. For smaller $\St$ the drift rate is limited by the terminal velocity of the dust, while for $\St>1$ the radial drift velocity is set by the torque from the drag forces, which is weaker for larger $\St$.}They also reported single particle velocity dispersions of order the sound speed, which is similar to what we find for the relative velocity of weakly coupled particles and therefore consistent with their motions being largely uncorrelated. They also noted that the motion of particle with $\St = 1$ were highly correlated, although they did not go as far as calculating the collision velocity between particles, which we find to be considerably smaller than the sound speed.

\subsection{2D vs. 3D}

The main uncertainties related to our work are due to the use of 2D simulations. For large $\St$, where the collision velocities are driven by gravitational forcing, the biggest effect will be due to the vertical component of the motion. Since the size of the spiral structures is of order a scale height in both the vertical and radial directions, the difference is unlikely to be more than a factor of two. The second uncertainty is due to the importance of small scale structure. However, since our simulations show a similar cut off in the azimuthal power spectrum of the \emph{density} fluctuations to the 3D simulations of \citet{Boley2007} we are confident that small scale structure will not be important when the driving is dominated by gravitational forcing.

Conversely, since we find that coupling to small scale fluctuations in the \emph{velocity} field is important for driving particles with $\St < 3$, the small scale structure \emph{will} be important for small particles. Using high resolution 2D simulations, \citet{Gibbons2015} found that self-gravity drives short-lived eddies with a minimum scale of $\sim 0.1H$, which corresponds to $\lambda_{\rm stop}$ for $\St = 1$. Therefore, we expect that in 2D even driving by gas drag is unlikely to drive strong collisions for $\St < 0.3$ (i.e. in the regime where we are unable to measure the dust velocity dispersion accurately in our simulations). However, it is currently unknown whether these results will extend to 3D or whether the formation and destruction of eddies can drive fully-developed turbulence. If this is the case, we would expect that at small $\St$ collisions between equal sized dust particles in self-gravitating discs would be analogous to the turbulent case, in which the collision velocity also has an $\St^{1/2}$ dependence \citep{Ormel2007}. However, in either case we expect that the relative drift between particles of different sizes will be the dominant cause of collisions for $\St < 1$

Note that we do not predict strong settling of grains in self-gravitating discs even though we predict  that the \emph{local} velocity dispersion of small grains is very low. As found by the 3D simulations of \citet{Rice2004}, it turns out that the thickness of the dust layer is  comparable to the gas layer, independent of particle size. The reason for this is that large particles have $\vmed{} \sim c_s$, giving a particle layer thickness $H_p = \vmed{} \Omega^{-1} \sim H$. For small grains the scale-height is controlled by the balance of sedimentation and diffusion \citep{Dubrulle1995,Youdin2007} and the scale-height is given by ${H_p/ H \sim \sqrt{\alpha_z/\St}/ \sqrt{1 + \alpha_z/\St}}$, where $\alpha_z$ is the turbulent parameter for diffusion. While for small grains $\vmed{} \ll c_s$ on scales of $\lambda_{\rm stop}$, the large scale turbulent velocity of the gas is of order $c_s$ \citep{Shi2014}, suggesting $\alpha_z \sim 1$ and $H_p \sim H$ for $\St \ll 1$. Therefore, we expect $H_p \sim H$ to hold more or less independently of particle size. With such large scale-heights, the self-gravity in the dust layer is likely to be unimportant unless growth to $\St \sim 0.3$ occurs.

\subsection{Grain growth to $\St > 0.3$}
\label{Sec:GrowthSt0.3}

\begin{figure*}
\begin{tabular}{cc}
\includegraphics[width=\columnwidth]{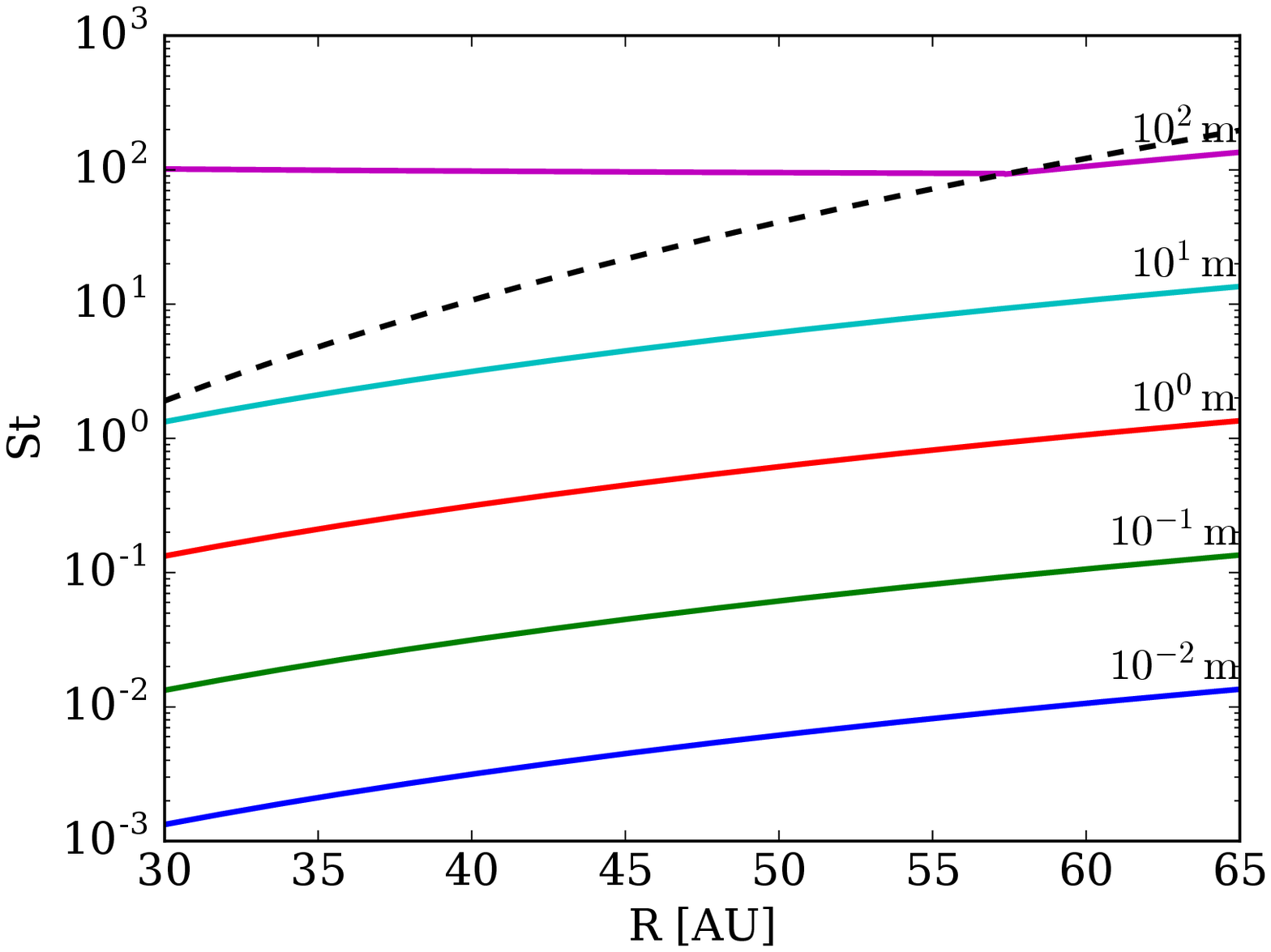} & \includegraphics[width=\columnwidth]{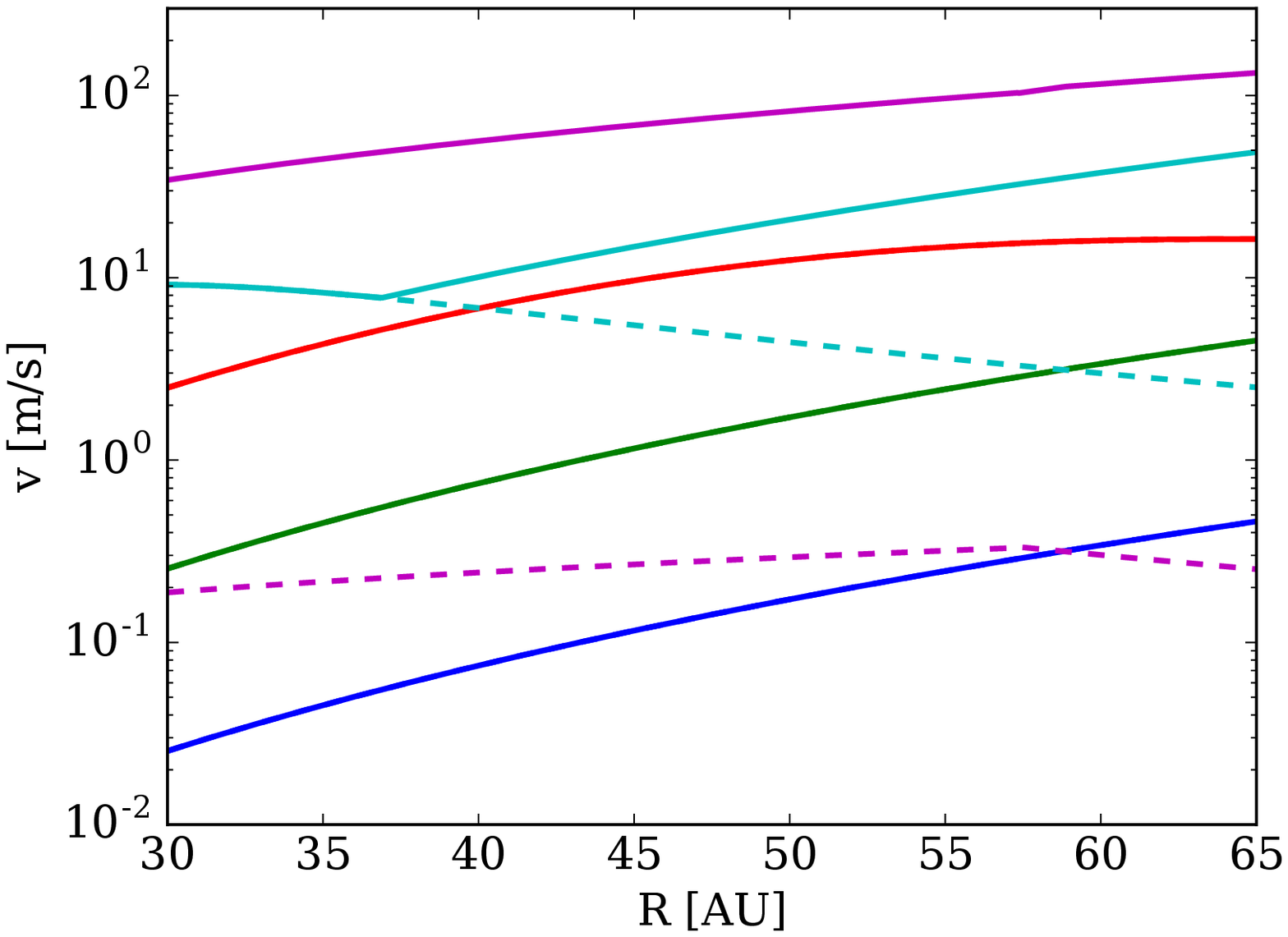} 
\end{tabular}
\caption{\emph{Left}: Stokes number as a function of radius for differing particle sizes adopting the analytic self-gravitating disc model of \citet{Clarke2009a}. The black dashed line shows the boundary between the Epstein and Stokes drag regimes. \emph{Right}: Corresponding collision velocity in a self-gravitating disc. Colour coding is as in the left panel.  The solid lines show  collision velocities calculated as being the maximum of the differential radial drift velocity (assuming particles differ by a factor of two in size) and the mono-disperse local relative velocity dispersion. In cases where the latter dominates, the former is shown by
the dashed lines. The transition to fully quadratic Stokes drag occurs for size above $100\unit{m}$. The velocities shown in the right hand panel are obtained from radial scalings of the simulation results (which equate to conditions at 60 AU in this disc model), being
derived from \autoref{Fig:MeanVel_St} for $100 >/  St > 1$, \autoref{Fig:Vel_bidisp} for $\St = 1$ and an extrapolation based on the short friction limit for smaller Stokes numbers. For $\St > 1$, the scaling of relative velocity with cooling timescale ($\beta$) is taken from the results of \autoref{Fig:MeanVel_Beta}.}
\label{Fig:Scalings}
\end{figure*}

We now consider whether it is possible, starting with arbitrarily small  grains,  for these  to grow to the point where direct gravitational collapse into planetesimals is possible. Such an outcome requires that the dust to gas ratio is significantly enhanced locally due to trapping of dust in spiral features. Efficient dust trapping requires two conditions to be met a) grains grow to a Stokes number of $\sim 0.3$ or above (this paper and \citealt{Gibbons2012}) and b) the spiral features are sufficiently pronounced that convergent radial flows within spiral features exceeds the dust's mean radial inflow\footnote{The spiral features must create pressure maxima. If the fluctuations are small enough then they can result in pressure changes relative to the background gradient without creating maxima}. In practice this implies that $\beta $ (the local ratio of disc cooling time to dynamical time) is sufficiently small ($ < 150 $; see \autoref{Sec:BiDisp} and \citealt{Gibbons2012}). The latter condition restricts the region of the disc in which this process can work (favouring larger radii); the former places a requirement that grains can grow to $St=0.3$ without experiencing collision velocities in excess of their fragmentation thresholds.

In our following discussion we  make several assumptions: i) For small grains the relative collision velocity is dominated by the differential radial drift velocity  acquired once the grain population achieves a modest (factor two) spread in sizes.  For small grains we expect that the scaling can be derived using the short friction time limit $\Delta v_{\rm drift} \sim (t_{s,1} - t_{s,2}) \nabla P / \rho$.  Since $\nabla P / \rho \sim (\Delta \Sigma / \Sigma H) c_s^2$ 
we expect $\Delta v_{\rm drift} \sim (\Delta \Sigma / \Sigma) (c_s^2/H) (t_{s,1} - t_{s,2})$, where $\Delta \Sigma/\Sigma$ is the amplitude of surface density fluctuations. ii) Drag is in the Epstein limit so that for grain size $a$, $t_s \propto a/\rho c_s$. iii) In the absence of observationally calibrated models for self-gravitating discs we use the pseudo-viscous models of self-regulated self-gravitating discs from \citet{Clarke2009a}. In particular, the simulations run here (for which  $\beta = 10$ and $H/R \sim 0.05$) correspond to a radius of $60\unit{au}$ and accretion rate (driven by gravitational torques in the disc) of $\sim 10^{-7}\,M_\odot\unit{yr}^{-1}$, with a sound speed of $\sim 200\unit{m\,s}^{-1}$. The disc in this region is optically thick with opacity dominated by ice grains. The relevant radial scalings for such a disc are: $c_s \propto R^{-1.5}$, $\rho \propto R^{-3}$ and $\Delta \Sigma/\Sigma \propto \beta^{-0.5} \propto R^{9/4}$. This
regime extends inwards to the point where ice grains sublime (around $15\unit{au}$ for this accretion rate). A radius of $30\unit{au}$ corresponds to $\beta = 150$
which we estimate as being the inner limit for which trapping in spirals is effective, given grains of a suitable Stokes number. \autoref{Fig:Scalings} illustrates the radial dependence of Stokes number and collision velocity as  a function of particle size given our model disc (see Figure caption for details).

Answering the question posed above then boils down to determining whether collision velocities are small enough to permit growth to St$=0.3$ at radii
$> 30\unit{au}$ (noting that at such radii the grains are expected to be icy)\footnote{This argument neglects the possibility of lucky growth, in which a small fraction of dust particles avoid high velocity collisions and grow beyond this size \citep{Windmark2012,Garaud2013}. However, unless runaway growth can occur this is unlikely to affect the bulk of the mass distribution. Also, current estimates of the effectiveness of lucky growth are model dependedent and likely sensitive to the collision velocity distribution used \citep{Drazkowska2014,Pan2014b}.}. If we fold together the radial scalings given above and the expression for
the differential radial drift of dust in multi-size dust populations, we find that the differential drift velocity between two particles with fixed sizes scales as:
\begin{equation}
\Delta v_{\rm drift} \propto R^{15/4}, \label{Eqn:vdrift}
\end{equation}
with radius since it is associated with a fixed Stokes number and, for our assumed disc structure, $St \propto R^3$. For typical parameters associated with self-gravitating discs, we find a maximum grain size 
\begin{equation}
s_{\rm max} \approx 
3 \left( \frac{v_{\rm frag}}{1\unit{m\,s}^{-1}}\right) 
\left( \frac{\rho_s}{1\unit{g\,cm}^{-3}}\right)^{-1}
\left( \frac{R}{60\unit{au}}\right)^{-15/4} \label{Eqn:MaxSize}
\unit{cm},
\end{equation}
and the corresponding maximum Stokes number is 
\begin{align}
&\St_{\max} \approx \nonumber \\
&~ 0.03
\left( \frac{v_{\rm frag}}{1\unit{m\,s}^{-1}}\right) 
\left( \frac{\rho_s}{1\unit{g\,cm}^{-3}}\right)^{-1}
\left( \frac{H}{5\unit{au}}\right)^{-1}
\left( \frac{R}{60\unit{au}}\right)^{-3/4}.
\end{align}

Since the fragmentation threshold for ices may be as high as $v_{\rm frag} > 15 \unit{m\,s}^{-1}$ \citep{Gundlach2015, Wada2009}, the growth of icy grains into the $\St = 0.3$ to $\St = 3$ range may be possible. Collisional growth much beyond these sizes is clearly impossible due to the high collision velocities, a fact that may actually help a significant fraction of the mass in icy dust to be concentrated in these critical sizes. However, detailed modelling is needed to determine whether growth to the critical sizes can occur since even a moderate fraction of collisions leading to fragmentation may prevent growth to the critical, metre scale, sizes as the net growth times may be long. Even so, gravitational fragmentation of icy grains may be a viable way to produce planetesimals at radii $>30\unit{au}$ early in the disc's evolution. 

If sufficient growth to $\St = 1$ does occur so that dust driven fragmentation takes place, then the typical mass of bound clumps should be the Jeans mass in the dust layer:
\begin{equation}
M_J = \frac{1}{6} \pi \rho_d \left(\frac{\pi \sigma_v^2}{G \rho_d}\right)^{3/2}.
\end{equation}
noting that $\sigma_v$ is the local velocity {\it dispersion} which, at low St, may be  much less than the differential radial drift velocity
(\autoref{Eqn:vdrift}). At $60\unit{au}$, where $\beta = 10$ and $\sigma_v \sim 0.1 c_s$ for $\St = 1$, we find  $M_J \approx 5 \times 10^{-2} (\rho_d/\rho_g)^{-1/2} M_\oplus$. For comparison, using shearing-sheet simulations \citet{Gibbons2014} found clump masses of order $3 \times 10^{-3} M_\oplus$ (for $\beta = 10$ and similar disc properties). While the simulations produce fragments that are $\sim$ a factor ten lower in mass than our estimates, this is to be expected since the simulations include the self-gravity of the dust.

Adopting the $\beta$ dependence of $\sigma_v$ from \autoref{Sec:BetaDep} and the radial scaling of $\beta$ expected in the regime of ice dominated cooling we obtain: 
\begin{equation}
\sigma_v \sim  \beta^{-0.5} v_k \St^{1/2} \propto \St^{1/2} R^{7/4}. \label{Eqn:v_disp}
\end{equation}
At $\St = 1$ this results in $\lambda_{\rm stop} < H$ for $R < 60\unit{au}$ and thus driving by self-gravity \emph{density} fluctuations will be inefficient, making \autoref{Eqn:v_disp} an over-estimate of the velocity dispersion. Instead we expect that the dominant driving of velocity \emph{dispersion} at smaller radii and $\St \sim 1$ may be the turbulent gas velocity, which is also weaker at longer cooling times \citep{Shi2014}. Therefore, we expect that the typical masses should be less than those estimated from \autoref{Eqn:v_disp} (${\approx 4.5 \times 10^{-5} (\rho_d/\rho_g)^{-1/2} M_\oplus}$ at $30\unit{au}$), with $10$ or $100\unit{km}$ planetesimals perhaps more typical outcomes than Moon-mass or Ceres-mass objects. 

Finally we note that even though we have shown that growth to $\St > 0.3$ (and hence significant particle concentration) requires {\it icy} grains, even silicate grains with $v_{\rm frag} \approx 1\unit{m\,s}^{-1}$ can grow to $\St \sim 10^{-2}$ or cm-sizes (\autoref{Eqn:MaxSize}). We thus expect spectral evidence of grain growth in self-gravitating discs even in the absence of icy grains.

While cm-sized grains correspond to Stokes numbers of $\sim 10^{-2}$ during the self-gravitating phase they will have $\St \gtrsim 1$ in typical Class II discs, once the accretion rate onto star has decreased and the correponding surface density has dropped. Although this increase in Stokes number means that perturbations driven by the disc are not as strongly damped, we have to also consider how the perturbations and corresponding collision velocity change as the disc evolves. To do this we once again make use of the pseudo-viscous models for self-gravitating discs and consider how the collision velocity changes as the accretion rate, $\dot{M}$, decreases towards the end of the self-gravitating phase.

\citet{Clarke2009a} showed that, while both $\Sigma \propto \dot{M}^{1/3}$ and $c_s \propto \dot{M}^{1/3}$, the cooling time $\beta$ does not depend on $\dot{M}$ (in the range of interest, $30$--$60\unit{au}$). Thus, in the Epstein regime, $\St \propto \dot{M}^{-1/3}$. Incorporating these behaviours into our scaling relations results in ${\sigma_v \propto \dot{M}^{-1/6}}$ (since ${\sigma_v \propto \St^{1/2}}$); $\Delta v_{\rm drift}$ is however independent of $\dot{M}$ since although $\St$ increases with decreasing $\dot{M}$, this is counteracted by the decrease in $c_s$. These relations show that the cm-size silicate grains will survive the self-gravitating phase and thus may already be present at the start of the Class II phase. Thus the detection of large grains (cm-size) in very young systems such as HL Tau and WL 12 may be evidence of effective grain growth during, or before, the self-gravitating phase.

\subsection{Weakly coupled particles: $\St > 3$}

We have seen above that we do not expect solids to grow into the weakly coupled `planetesimal' regime with $St > 3$ by coagulation alone, owing to
the rise in velocity dispersion with Stokes number. Nevertheless, we have argued that it may be possible to grow in this way to Stokes numbers ($0.3-3$) where grains are sufficiently concentrated so as to lead to gravitational collapse. In this way it is possible that self-gravitating discs may from a population of planetesimals as originally suggested by \citet{Rice2004}. We now consider the implications of our calculations for the subsequent evolution of such a planetesimal population.

For large grains and planetesimals with $\St \gg 1$ we found $\vmed{} \approx 2 c_s$, in good agreement with \citet{Walmswell2013}. For similar sized planetesimals, collisions at these velocities will be above the catastrophic disruption threshold, $Q^*_{\rm RD}$, except for the largest planetesimals for which their gravitational binding energy exceeds the collision kinetic energy ($s \gtrsim 100\unit{km}$, \citealt{Stewart2009}). However, collisions will be rare since the time between collisions, $t_c \sim 10^{4} (\tfrac{s}{\unit{m}}) \unit{yr}$, exceeds the lifetime of the self-gravitating phase (a few $10^5\unit{yr}$) for sizes above a few 10s of meters (the above estimate makes the optimistic assumption that the entire solid component of the disc is in the form of objects of size $s$).

We can also use \autoref{Fig:Scalings} to examine typical velocity dispersions in the planetesimal regime. This plot takes into account the fact that the form of
the drag term transitions from the Epstein regime to the `Stokes drag' regime in which the drag varies quadratically with relative velocity and hence the
Stokes number is a function of relative velocity as well as particle size and disc parameters. This transition occurs for particles with size $s > 9 \lambda_{\rm mfp}/4$, where $\lambda_{\rm mfp}$ is the mean free path, the drag forces transitions to the Stokes regime, in which $\St$ is a function of velocity as well as size \citep{Whipple1972,Weidenschilling1977}. Since $\lambda_{\rm mfp} \approx 50 (R/60 \unit{au})^3 \unit{m}$, particles with $\St < 1$ are in the Epstein regime, but for particles with $\St > 1$ we have to consider the transition to Stokes drag, which happens at
\begin{equation}
s \approx 110 \left(\frac{R}{60 \unit{au}} \right)^{3} \unit{m},
\end{equation}
which corresponds to $\St \approx 120 (R/60\unit{au})^6$. The transition to the fully quadratic Stokes drag regime occurs at $s \approx 500 (R/60\unit{au})^2 \unit{m}$.

\autoref{Fig:Scalings} shows that planestesimal scale objects are generally in the Stokes drag regime and that the velocity dispersion falls steeply at small radii, a result that is readily explicable in terms of the normalised cooling timescale ($\beta$) dependence shown in \autoref{Fig:MeanVel_Beta} and the fact that the cooling time is long at small radii. This suggests that the eccentricity of planetesimals will be predominantly driven at apocentre as argued by \citet{Walmswell2013}. Thus we expect that if a population of planetesimals is formed by this mechanism at radii $> 30\unit{au}$ (as argued in \ref{Sec:GrowthSt0.3}) then they will be perturbed into eccentric orbits which visit the inner disc but which retain apocentres $> 30\unit{au}$. Using the numerical results of \citet{Walmswell2013}, we find that few planetesimals are likely to end up, at the end of the disc's self-gravitating phase, at radii well within $30\unit{au}$: this is because this represents a large radial excursion for any planetesimals formed in the outer part of the disc (where spiral structure is strongest) while for those planetesimals formed at the limiting radius of $30\unit{au}$ the relatively weak spiral structure results in weak orbital excitation.

Even though we expect the radial re-configuration of the planetesimal distribution to be minor, re-scaling the results of Walmswell et al suggests that the planetesimals would nevertheless end up on rather eccentric orbits (typically in the range $e \sim 0.1-0.5$) over the lifetime of the self-gravitating disc. This value should be contrasted with the typical eccentricities of planetesimals established through equilibrium between gas drag and stirring by larger bodies in \emph{non}-self gravitating discs ($\sim 0.01$; \citealt{Kukobu2000}). The high eccentricities of any planetesimals formed in the self-gravitating phase means that
they would {\it not} be significantly damped by gas drag over the entire age of the gas disc, since the estimated eccentricity damping timescale given these initial
eccentricities is $> 10^8$ years \citep{Ida2008}. Similarly, the high velocity dispersion means that gravitational focussing is ineffective in enhancing the rate of mutual collisions, so that typical collision times are in excess of $10^7$ years. Therefore, we can expect the population of eccentric planetesimals to be long lived unless disturbed by some other process, such as planetary migration. 

\section{Conclusion}
\label{Sec:Conclude}

Our simulations show that the spiral structure in self-gravitating protoplanetary discs is only effective at driving a large velocity dispersion in the dust for particles with  Stokes number $St$ (ratio of stopping time to dynamical time) $\St \gtrsim 3$. Below $\St \sim 1$ the velocity dispersion is driven by coupling to small scale structure in the gas velocity via drag forces,  resulting in collision velocities that are considerably less than the sound speed. Instead, the differing drag forces felt by particles of different sizes gives rise to different terminal velocities larger than the velocity dispersion for particles of a single size.

The relative velocities produced by differential radial drift in multi-sized grain populations are also an important consideration for particle growth in non-self gravitating discs \citep{Dullemond2005,Brauer2008} and in the inner regions of self-gravitating discs where the spiral structure is weak. Here, however, we concentrate on the regime where spiral structure is strong enough to produce significant grain trapping in density maxima and where high local values of the solid to gas ratio can trigger collective phenomena (ultimately gravitational collapse) of the solid component. The regime of strong spiral structure, where there is the potential for trapping and concentration of solids, corresponds to radii in the disc beyond $\sim 30 \unit{au}$.

Efficient trapping however requires, in addition to regions of strong spiral structure, that particles can grow so as to enter a critical regime of Stokes numbers: $0.3 < St < 3$. Thus growth to these sizes ($\St \sim 1$ corresponds to metre sizes at $60\unit{au}$; see left panel of \autoref{Fig:Scalings}).  We
find that the median relative velocity of metre sized objects at $60\unit{au}$ is around $10\unit{m\,s}^{-1}$ (see right hand panel of \autoref{Fig:Scalings}). This clearly rules out sufficient growth in the case of silicate dominated grains (for which the fragmentation threshold is around $1\unit{m\,s}^{-1}$) whose growth would be limited to cm-sizes. However we note that growth to cm scales is sufficient to produce observational signatures of grain growth in self-gravitating discs.

The best prospects for direct gravitational collapse comes from icy grains since the fragmentation threshold of ices is above $15\unit{m\,s}^{-1}$. We noted above that efficient dust trapping is restricted to regions outside $30$ A.U., where the spiral structure is sufficiently strong. Provided that $\dot{M} < 10^{-6}\,M_\odot\unit{yr}^{-1}$ the water snow line corresponds to $<30\unit{au}$. We thus argue that icy solids may be able to grow to the point where they can be concentrated and ultimately undergo gravitational collapse in this region of the disc.

 If gravitational collapse indeed occurs, then the size scale of fragments (`planetesimals') is largely controlled by the driving  of the dust velocity by turbulent gas fluctuations. Although the resulting planetesimals will have their eccentricities pumped by interaction with spiral structure in the disc, they will remain largely confined to the outer disc. Their high eccentricities imply very long collision times and very long timescales for drag from the disc gas,
so that they are unlikely to be further modified during the lifetime of the gas disc. Therefore we suggest that the legacy of particle growth during the earliest, self-gravitating, phase of the disc's evolution is likely to be a population of icy planetesimals on eccentric orbits in the outer disc.

\section*{Acknowledgements}

We thank Phil Armitage, Alex Hubbard and Andrew Shannon for useful discussions and insights. We thank the anonymous referee for a careful reading of the manuscript and constructive comments that helped significantly improve the manuscript. This work has been supported by the DISCSIM project, grant agreement 341137 funded by the European Research Council under ERC-2013-ADG and has used the DIRAC Shared Memory Processing and DiRAC Data Analytic systems at the University of Cambridge. The DIRAC Shared Memory Processing system is operated by the COSMOS Project at the Department of Applied Mathematics and Theoretical Physics and was funded by BIS National E-infrastructure capital grant ST/J005673/1, STFC capital grant ST/H008586/1. The DiRAC Data Analytic system was funded by BIS National E-infrastructure capital grant ST/J005673/1, STFC capital grant ST/H008586/1. Both systems are on behalf of the STFC DiRAC HPC Facility (www.dirac.ac.uk), funded by the STFC DiRAC Operations grant ST/K00333X/1. 

\footnotesize{
    \bibliography{dust}

\begin{thebibliography}{}
\makeatletter
\relax
\def\mn@urlcharsother{\let\do\@makeother \do\$\do\&\do\#\do\^\do\_\do\%\do\~}
\def\mn@doi{\begingroup\mn@urlcharsother \@ifnextchar [ {\mn@doi@}
  {\mn@doi@[]}}
\def\mn@doi@[#1]#2{\def\@tempa{#1}\ifx\@tempa\@empty \href
  {http://dx.doi.org/#2} {doi:#2}\else \href {http://dx.doi.org/#2} {#1}\fi
  \endgroup}
\def\mn@eprint#1#2{\mn@eprint@#1:#2::\@nil}
\def\mn@eprint@arXiv#1{\href {http://arxiv.org/abs/#1} {{\tt arXiv:#1}}}
\def\mn@eprint@dblp#1{\href {http://dblp.uni-trier.de/rec/bibtex/#1.xml}
  {dblp:#1}}
\def\mn@eprint@#1:#2:#3:#4\@nil{\def\@tempa {#1}\def\@tempb {#2}\def\@tempc
  {#3}\ifx \@tempc \@empty \let \@tempc \@tempb \let \@tempb \@tempa \fi \ifx
  \@tempb \@empty \def\@tempb {arXiv}\fi \@ifundefined
  {mn@eprint@\@tempb}{\@tempb:\@tempc}{\expandafter \expandafter \csname
  mn@eprint@\@tempb\endcsname \expandafter{\@tempc}}}

\bibitem[\protect\citeauthoryear{{ALMA Partnership} et~al.,}{{ALMA Partnership}
  et~al.}{2015}]{ALMA2015}
{ALMA Partnership} et~al., 2015, \mn@doi [\apjl] {10.1088/2041-8205/808/1/L3},
  \href{http://adsabs.harvard.edu/abs/2015ApJ...808L...3A}{808, L3}

\bibitem[\protect\citeauthoryear{{Bai} \& {Stone}}{{Bai} \&
  {Stone}}{2010a}]{Bai2010a}
{Bai} X.-N.,  {Stone} J.~M.,  2010a, \mn@doi [\apj]
  {10.1088/0004-637X/722/2/1437},
  \href{http://ukads.nottingham.ac.uk/abs/2010ApJ...722.1437B}{722, 1437}

\bibitem[\protect\citeauthoryear{{Bai} \& {Stone}}{{Bai} \&
  {Stone}}{2010b}]{Bai2010b}
{Bai} X.-N.,  {Stone} J.~M.,  2010b, \mn@doi [\apjl]
  {10.1088/2041-8205/722/2/L220},
  \href{http://ukads.nottingham.ac.uk/abs/2010ApJ...722L.220B}{722, L220}

\bibitem[\protect\citeauthoryear{{Barnes} \& {Hut}}{{Barnes} \&
  {Hut}}{1986}]{Barnes1986}
{Barnes} J.,  {Hut} P.,  1986, \mn@doi [\nat] {10.1038/324446a0},
  \href{http://ukads.nottingham.ac.uk/abs/1986Natur.324..446B}{324, 446}

\bibitem[\protect\citeauthoryear{{Boley}, {Durisen}, {Nordlund}  \&
  {Lord}}{{Boley} et~al.}{2007}]{Boley2007}
{Boley} A.~C.,  {Durisen} R.~H.,  {Nordlund} {\AA}.,   {Lord} J.,  2007,
  \mn@doi [\apj] {10.1086/519767},
  \href{http://ukads.nottingham.ac.uk/abs/2007ApJ...665.1254B}{665, 1254}

\bibitem[\protect\citeauthoryear{{Booth}, {Sijacki}  \& {Clarke}}{{Booth}
  et~al.}{2015}]{Booth2015}
{Booth} R.~A.,  {Sijacki} D.,   {Clarke} C.~J.,  2015, \mn@doi [\mnras]
  {10.1093/mnras/stv1486},
  \href{http://ukads.nottingham.ac.uk/abs/2015MNRAS.452.3932B}{452, 3932}

\bibitem[\protect\citeauthoryear{{Brauer}, {Henning}  \& {Dullemond}}{{Brauer}
  et~al.}{2008}]{Brauer2008}
{Brauer} F.,  {Henning} T.,   {Dullemond} C.~P.,  2008, \mn@doi [\aap]
  {10.1051/0004-6361:200809780},
  \href{http://ukads.nottingham.ac.uk/abs/2008A%26A...487L...1B}{487, L1}

\bibitem[\protect\citeauthoryear{{Carballido}, {Cuzzi}  \&
  {Hogan}}{{Carballido} et~al.}{2010}]{Carballido2010}
{Carballido} A.,  {Cuzzi} J.~N.,   {Hogan} R.~C.,  2010, \mn@doi [\mnras]
  {10.1111/j.1365-2966.2010.16653.x},
  \href{http://adsabs.harvard.edu/abs/2010MNRAS.405.2339C}{405, 2339}

\bibitem[\protect\citeauthoryear{{Carrera}, {Johansen}  \& {Davies}}{{Carrera}
  et~al.}{2015}]{Carrera2015}
{Carrera} D.,  {Johansen} A.,   {Davies} M.~B.,  2015, \mn@doi [\aap]
  {10.1051/0004-6361/201425120},
  \href{http://ukads.nottingham.ac.uk/abs/2015A%26A...579A..43C}{579, A43}

\bibitem[\protect\citeauthoryear{{Clarke}}{{Clarke}}{2009}]{Clarke2009a}
{Clarke} C.~J.,  2009, \mn@doi [\mnras] {10.1111/j.1365-2966.2009.14774.x},
  \href{http://ukads.nottingham.ac.uk/abs/2009MNRAS.396.1066C}{396, 1066}

\bibitem[\protect\citeauthoryear{{Clarke} \& {Lodato}}{{Clarke} \&
  {Lodato}}{2009}]{Clarke2009}
{Clarke} C.~J.,  {Lodato} G.,  2009, \mn@doi [\mnras]
  {10.1111/j.1745-3933.2009.00695.x},
  \href{http://ukads.nottingham.ac.uk/abs/2009MNRAS.398L...6C}{398, L6}

\bibitem[\protect\citeauthoryear{{Cossins}, {Lodato}  \& {Clarke}}{{Cossins}
  et~al.}{2009}]{Cossins2009}
{Cossins} P.,  {Lodato} G.,   {Clarke} C.~J.,  2009, \mn@doi [\mnras]
  {10.1111/j.1365-2966.2008.14275.x},
  \href{http://adsabs.harvard.edu/abs/2009MNRAS.393.1157C}{393, 1157}

\bibitem[\protect\citeauthoryear{{Cullen} \& {Dehnen}}{{Cullen} \&
  {Dehnen}}{2010}]{Cullen2010}
{Cullen} L.,  {Dehnen} W.,  2010, \mn@doi [\mnras]
  {10.1111/j.1365-2966.2010.17158.x},
  \href{http://ukads.nottingham.ac.uk/abs/2010MNRAS.408..669C}{408, 669}

\bibitem[\protect\citeauthoryear{{Cuzzi} \& {Hogan}}{{Cuzzi} \&
  {Hogan}}{2003}]{Cuzzi2003}
{Cuzzi} J.~N.,  {Hogan} R.~C.,  2003, \mn@doi [\icarus]
  {10.1016/S0019-1035(03)00104-0},
  \href{http://ukads.nottingham.ac.uk/abs/2003Icar..164..127C}{164, 127}

\bibitem[\protect\citeauthoryear{{Dehnen}}{{Dehnen}}{2001}]{Dehnen2001}
{Dehnen} W.,  2001, \mn@doi [\mnras] {10.1046/j.1365-8711.2001.04237.x},
  \href{http://ukads.nottingham.ac.uk/abs/2001MNRAS.324..273D}{324, 273}

\bibitem[\protect\citeauthoryear{{Dehnen} \& {Aly}}{{Dehnen} \&
  {Aly}}{2012}]{Dehnen2012}
{Dehnen} W.,  {Aly} H.,  2012, \mn@doi [\mnras]
  {10.1111/j.1365-2966.2012.21439.x},
  \href{http://ukads.nottingham.ac.uk/abs/2012MNRAS.425.1068D}{425, 1068}

\bibitem[\protect\citeauthoryear{{Dr{\c a}{\.z}kowska}, {Windmark}  \&
  {Dullemond}}{{Dr{\c a}{\.z}kowska} et~al.}{2014}]{Drazkowska2014}
{Dr{\c a}{\.z}kowska} J.,  {Windmark} F.,   {Dullemond} C.~P.,  2014, \mn@doi
  [\aap] {10.1051/0004-6361/201423708},
  \href{http://ukads.nottingham.ac.uk/abs/2014A%26A...567A..38D}{567, A38}

\bibitem[\protect\citeauthoryear{{Dubrulle}, {Morfill}  \&
  {Sterzik}}{{Dubrulle} et~al.}{1995}]{Dubrulle1995}
{Dubrulle} B.,  {Morfill} G.,   {Sterzik} M.,  1995, \mn@doi [\icarus]
  {10.1006/icar.1995.1058},
  \href{http://ukads.nottingham.ac.uk/abs/1995Icar..114..237D}{114, 237}

\bibitem[\protect\citeauthoryear{{Dullemond} \& {Dominik}}{{Dullemond} \&
  {Dominik}}{2005}]{Dullemond2005}
{Dullemond} C.~P.,  {Dominik} C.,  2005, \mn@doi [\aap]
  {10.1051/0004-6361:20042080},
  \href{http://ukads.nottingham.ac.uk/abs/2005A%26A...434..971D}{434, 971}

\bibitem[\protect\citeauthoryear{{Eisner}}{{Eisner}}{2012}]{Eisner2012}
{Eisner} J.~A.,  2012, \mn@doi [\apj] {10.1088/0004-637X/755/1/23},
  \href{http://ukads.nottingham.ac.uk/abs/2012ApJ...755...23E}{755, 23}

\bibitem[\protect\citeauthoryear{{Falkovich}, {Fouxon}  \&
  {Stepanov}}{{Falkovich} et~al.}{2002}]{Falkovich2002}
{Falkovich} G.,  {Fouxon} A.,   {Stepanov} M.~G.,  2002, \mn@doi [\nat]
  {10.1038/nature00983},
  \href{http://ukads.nottingham.ac.uk/abs/2002Natur.419..151F}{419, 151}

\bibitem[\protect\citeauthoryear{{Garaud}, {Meru}, {Galvagni}  \&
  {Olczak}}{{Garaud} et~al.}{2013}]{Garaud2013}
{Garaud} P.,  {Meru} F.,  {Galvagni} M.,   {Olczak} C.,  2013, \mn@doi [\apj]
  {10.1088/0004-637X/764/2/146},
  \href{http://ukads.nottingham.ac.uk/abs/2013ApJ...764..146G}{764, 146}

\bibitem[\protect\citeauthoryear{{Garc{\'{\i}}a-Senz}, {Cabez{\'o}n}  \&
  {Escart{\'{\i}}n}}{{Garc{\'{\i}}a-Senz} et~al.}{2012}]{Garcia-Senz2012}
{Garc{\'{\i}}a-Senz} D.,  {Cabez{\'o}n} R.~M.,   {Escart{\'{\i}}n} J.~A.,
  2012, \mn@doi [\aap] {10.1051/0004-6361/201117939},
  \href{http://ukads.nottingham.ac.uk/abs/2012A&A...538A...9G}{538, A9}

\bibitem[\protect\citeauthoryear{{Gibbons}, {Rice}  \&
  {Mamatsashvili}}{{Gibbons} et~al.}{2012}]{Gibbons2012}
{Gibbons} P.~G.,  {Rice} W.~K.~M.,   {Mamatsashvili} G.~R.,  2012, \mn@doi
  [\mnras] {10.1111/j.1365-2966.2012.21731.x},
  \href{http://ukads.nottingham.ac.uk/abs/2012MNRAS.426.1444G}{426, 1444}

\bibitem[\protect\citeauthoryear{{Gibbons}, {Mamatsashvili}  \&
  {Rice}}{{Gibbons} et~al.}{2014}]{Gibbons2014}
{Gibbons} P.~G.,  {Mamatsashvili} G.~R.,   {Rice} W.~K.~M.,  2014, \mn@doi
  [\mnras] {10.1093/mnras/stu809},
  \href{http://ukads.nottingham.ac.uk/abs/2014MNRAS.442..361G}{442, 361}

\bibitem[\protect\citeauthoryear{{Gibbons}, {Mamatsashvili}  \&
  {Rice}}{{Gibbons} et~al.}{2015}]{Gibbons2015}
{Gibbons} P.~G.,  {Mamatsashvili} G.~R.,   {Rice} W.~K.~M.,  2015, \mn@doi
  [\mnras] {10.1093/mnras/stv1766},
  \href{http://ukads.nottingham.ac.uk/abs/2015MNRAS.453.4232G}{453, 4232}

\bibitem[\protect\citeauthoryear{{Greaves}, {Richards}, {Rice}  \&
  {Muxlow}}{{Greaves} et~al.}{2008}]{Greaves2008}
{Greaves} J.~S.,  {Richards} A.~M.~S.,  {Rice} W.~K.~M.,   {Muxlow} T.~W.~B.,
  2008, \mn@doi [\mnras] {10.1111/j.1745-3933.2008.00559.x},
  \href{http://ukads.nottingham.ac.uk/abs/2008MNRAS.391L..74G}{391, L74}

\bibitem[\protect\citeauthoryear{{Gundlach} \& {Blum}}{{Gundlach} \&
  {Blum}}{2015}]{Gundlach2015}
{Gundlach} B.,  {Blum} J.,  2015, \mn@doi [\apj] {10.1088/0004-637X/798/1/34},
  \href{http://ukads.nottingham.ac.uk/abs/2015ApJ...798...34G}{798, 34}

\bibitem[\protect\citeauthoryear{{Gustavsson} \& {Mehlig}}{{Gustavsson} \&
  {Mehlig}}{2011}]{Gustavsson2011}
{Gustavsson} K.,  {Mehlig} B.,  2011, \mn@doi [\pre]
  {10.1103/PhysRevE.84.045304},
  \href{http://ukads.nottingham.ac.uk/abs/2011PhRvE..84d5304G}{84, 045304}

\bibitem[\protect\citeauthoryear{{Gustavsson}, {Mehlig}, {Wilkinson}  \&
  {Uski}}{{Gustavsson} et~al.}{2008}]{Gustavsson2008}
{Gustavsson} K.,  {Mehlig} B.,  {Wilkinson} M.,   {Uski} V.,  2008, \mn@doi
  [Physical Review Letters] {10.1103/PhysRevLett.101.174503},
  \href{http://ukads.nottingham.ac.uk/abs/2008PhRvL.101q4503G}{101, 174503}

\bibitem[\protect\citeauthoryear{{G{\"u}ttler}, {Blum}, {Zsom}, {Ormel}  \&
  {Dullemond}}{{G{\"u}ttler} et~al.}{2010}]{Guttler2010}
{G{\"u}ttler} C.,  {Blum} J.,  {Zsom} A.,  {Ormel} C.~W.,   {Dullemond} C.~P.,
  2010, \mn@doi [\aap] {10.1051/0004-6361/200912852},
  \href{http://ukads.nottingham.ac.uk/abs/2010A&A...513A..56G}{513, A56}

\bibitem[\protect\citeauthoryear{{Ida}, {Guillot}  \& {Morbidelli}}{{Ida}
  et~al.}{2008}]{Ida2008}
{Ida} S.,  {Guillot} T.,   {Morbidelli} A.,  2008, \mn@doi [\apj]
  {10.1086/591903},
  \href{http://ukads.nottingham.ac.uk/abs/2008ApJ...686.1292I}{686, 1292}

\bibitem[\protect\citeauthoryear{{Johansen}, {Oishi}, {Mac Low}, {Klahr},
  {Henning}  \& {Youdin}}{{Johansen} et~al.}{2007}]{Johansen2007}
{Johansen} A.,  {Oishi} J.~S.,  {Mac Low} M.-M.,  {Klahr} H.,  {Henning} T.,
  {Youdin} A.,  2007, \mn@doi [\nat] {10.1038/nature06086},
  \href{http://ukads.nottingham.ac.uk/abs/2007Natur.448.1022J}{448, 1022}

\bibitem[\protect\citeauthoryear{{Johansen}, {Blum}, {Tanaka}, {Ormel},
  {Bizzarro}  \& {Rickman}}{{Johansen} et~al.}{2014}]{Johansen2014}
{Johansen} A.,  {Blum} J.,  {Tanaka} H.,  {Ormel} C.,  {Bizzarro} M.,
  {Rickman} H.,  2014, \mn@doi [Protostars and Planets VI]
  {10.2458/azu_uapress_9780816531240-ch024},
  \href{http://ukads.nottingham.ac.uk/abs/2014prpl.conf..547J}{pp 547--570}

\bibitem[\protect\citeauthoryear{{Kokubo}, {Ida}  \& {Makino}}{{Kokubo}
  et~al.}{2000}]{Kukobu2000}
{Kokubo} E.,  {Ida} S.,   {Makino} J.,  2000, \mn@doi [\icarus]
  {10.1006/icar.2000.6496},
  \href{http://ukads.nottingham.ac.uk/abs/2000Icar..148..419K}{148, 419}

\bibitem[\protect\citeauthoryear{{Lambrechts} \& {Johansen}}{{Lambrechts} \&
  {Johansen}}{2012}]{Lambrechts2012}
{Lambrechts} M.,  {Johansen} A.,  2012, \mn@doi [\aap]
  {10.1051/0004-6361/201219127},
  \href{http://ukads.nottingham.ac.uk/abs/2012A%26A...544A..32L}{544, A32}

\bibitem[\protect\citeauthoryear{{Laughlin}, {Steinacker}  \&
  {Adams}}{{Laughlin} et~al.}{2004}]{Laughlin2004}
{Laughlin} G.,  {Steinacker} A.,   {Adams} F.~C.,  2004, \mn@doi [\apj]
  {10.1086/386316},
  \href{http://ukads.nottingham.ac.uk/abs/2004ApJ...608..489L}{608, 489}

\bibitem[\protect\citeauthoryear{{Lodato}}{{Lodato}}{2007}]{Lodato2007}
{Lodato} G.,  2007, \mn@doi [Nuovo Cimento Rivista Serie]
  {10.1393/ncr/i2007-10022-x},
  \href{http://ukads.nottingham.ac.uk/abs/2007NCimR..30..293L}{30, 293}

\bibitem[\protect\citeauthoryear{{Lodato} \& {Rice}}{{Lodato} \&
  {Rice}}{2004}]{Lodato2004}
{Lodato} G.,  {Rice} W.~K.~M.,  2004, \mn@doi [\mnras]
  {10.1111/j.1365-2966.2004.07811.x},
  \href{http://ukads.nottingham.ac.uk/abs/2004MNRAS.351..630L}{351, 630}

\bibitem[\protect\citeauthoryear{{Mann}, {Andrews}, {Eisner}, {Williams},
  {Meyer}, {Di Francesco}, {Carpenter}  \& {Johnstone}}{{Mann}
  et~al.}{2015}]{Mann2015}
{Mann} R.~K.,  {Andrews} S.~M.,  {Eisner} J.~A.,  {Williams} J.~P.,  {Meyer}
  M.~R.,  {Di Francesco} J.,  {Carpenter} J.~M.,   {Johnstone} D.,  2015,
  \mn@doi [\apj] {10.1088/0004-637X/802/2/77},
  \href{http://ukads.nottingham.ac.uk/abs/2015ApJ...802...77M}{802, 77}

\bibitem[\protect\citeauthoryear{{Meru} \& {Bate}}{{Meru} \&
  {Bate}}{2011}]{Meru2011}
{Meru} F.,  {Bate} M.~R.,  2011, \mn@doi [\mnras]
  {10.1111/j.1745-3933.2010.00978.x},
  \href{http://ukads.nottingham.ac.uk/abs/2011MNRAS.411L...1M}{411, L1}

\bibitem[\protect\citeauthoryear{{Meru} \& {Bate}}{{Meru} \&
  {Bate}}{2012}]{Meru2012}
{Meru} F.,  {Bate} M.~R.,  2012, \mn@doi [\mnras]
  {10.1111/j.1365-2966.2012.22035.x},
  \href{http://ukads.nottingham.ac.uk/abs/2012MNRAS.427.2022M}{427, 2022}

\bibitem[\protect\citeauthoryear{{Michael}, {Steiman-Cameron}, {Durisen}  \&
  {Boley}}{{Michael} et~al.}{2012}]{Michael2012}
{Michael} S.,  {Steiman-Cameron} T.~Y.,  {Durisen} R.~H.,   {Boley} A.~C.,
  2012, \mn@doi [\apj] {10.1088/0004-637X/746/1/98},
  \href{http://ukads.nottingham.ac.uk/abs/2012ApJ...746...98M}{746, 98}

\bibitem[\protect\citeauthoryear{{Miotello}, {Testi}, {Lodato}, {Ricci},
  {Rosotti}, {Brooks}, {Maury}  \& {Natta}}{{Miotello}
  et~al.}{2014}]{Miotello2014}
{Miotello} A.,  {Testi} L.,  {Lodato} G.,  {Ricci} L.,  {Rosotti} G.,  {Brooks}
  K.,  {Maury} A.,   {Natta} A.,  2014, \mn@doi [\aap]
  {10.1051/0004-6361/201322945},
  \href{http://ukads.nottingham.ac.uk/abs/2014A%26A...567A..32M}{567, A32}

\bibitem[\protect\citeauthoryear{{M{\"u}ller}, {Kley}  \& {Meru}}{{M{\"u}ller}
  et~al.}{2012}]{Muller2012}
{M{\"u}ller} T.~W.~A.,  {Kley} W.,   {Meru} F.,  2012, \mn@doi [\aap]
  {10.1051/0004-6361/201118737},
  \href{http://ukads.nottingham.ac.uk/abs/2012A&A...541A.123M}{541, A123}

\bibitem[\protect\citeauthoryear{{Nakagawa}, {Sekiya}  \& {Hayashi}}{{Nakagawa}
  et~al.}{1986}]{Nakagawa1986}
{Nakagawa} Y.,  {Sekiya} M.,   {Hayashi} C.,  1986, \mn@doi [\icarus]
  {10.1016/0019-1035(86)90121-1},
  \href{http://ukads.nottingham.ac.uk/abs/1986Icar...67..375N}{67, 375}

\bibitem[\protect\citeauthoryear{{Ormel} \& {Cuzzi}}{{Ormel} \&
  {Cuzzi}}{2007}]{Ormel2007}
{Ormel} C.~W.,  {Cuzzi} J.~N.,  2007, \mn@doi [\aap]
  {10.1051/0004-6361:20066899},
  \href{http://ukads.nottingham.ac.uk/abs/2007A&A...466..413O}{466, 413}

\bibitem[\protect\citeauthoryear{{Pan} \& {Padoan}}{{Pan} \&
  {Padoan}}{2010}]{Pan2010}
{Pan} L.,  {Padoan} P.,  2010, \mn@doi [Journal of Fluid Mechanics]
  {10.1017/S0022112010002855},
  \href{http://ukads.nottingham.ac.uk/abs/2010JFM...661...73P}{661, 73}

\bibitem[\protect\citeauthoryear{{Pan} \& {Padoan}}{{Pan} \&
  {Padoan}}{2013}]{Pan2013}
{Pan} L.,  {Padoan} P.,  2013, \mn@doi [\apj] {10.1088/0004-637X/776/1/12},
  \href{http://ukads.nottingham.ac.uk/abs/2013ApJ...776...12P}{776, 12}

\bibitem[\protect\citeauthoryear{{Pan} \& {Padoan}}{{Pan} \&
  {Padoan}}{2014}]{Pan2014b}
{Pan} L.,  {Padoan} P.,  2014, \mn@doi [\apj] {10.1088/0004-637X/797/2/101},
  \href{http://ukads.nottingham.ac.uk/abs/2014ApJ...797..101P}{797, 101}

\bibitem[\protect\citeauthoryear{{Pan}, {Padoan}  \& {Scalo}}{{Pan}
  et~al.}{2014}]{Pan2014a}
{Pan} L.,  {Padoan} P.,   {Scalo} J.,  2014, \mn@doi [\apj]
  {10.1088/0004-637X/791/1/48},
  \href{http://ukads.nottingham.ac.uk/abs/2014ApJ...791...48P}{791, 48}

\bibitem[\protect\citeauthoryear{{Pinilla}, {Birnstiel}, {Ricci}, {Dullemond},
  {Uribe}, {Testi}  \& {Natta}}{{Pinilla} et~al.}{2012}]{Pinilla2012}
{Pinilla} P.,  {Birnstiel} T.,  {Ricci} L.,  {Dullemond} C.~P.,  {Uribe} A.~L.,
   {Testi} L.,   {Natta} A.,  2012, \mn@doi [\aap]
  {10.1051/0004-6361/201118204},
  \href{http://ukads.nottingham.ac.uk/abs/2012A%26A...538A.114P}{538, A114}

\bibitem[\protect\citeauthoryear{{Price}}{{Price}}{2012}]{Price2012}
{Price} D.~J.,  2012, \mn@doi [Journal of Computational Physics]
  {10.1016/j.jcp.2010.12.011},
  \href{http://ukads.nottingham.ac.uk/abs/2012JCoPh.231..759P}{231, 759}

\bibitem[\protect\citeauthoryear{{Price} \& {Monaghan}}{{Price} \&
  {Monaghan}}{2007}]{Price2007}
{Price} D.~J.,  {Monaghan} J.~J.,  2007, \mn@doi [\mnras]
  {10.1111/j.1365-2966.2006.11241.x},
  \href{http://ukads.nottingham.ac.uk/abs/2007MNRAS.374.1347P}{374, 1347}

\bibitem[\protect\citeauthoryear{{Rafikov}}{{Rafikov}}{2005}]{Rafikov2005}
{Rafikov} R.~R.,  2005, \mn@doi [\apjl] {10.1086/428899},
  \href{http://ukads.nottingham.ac.uk/abs/2005ApJ...621L..69R}{621, L69}

\bibitem[\protect\citeauthoryear{{Ricci}, {Testi}, {Natta}  \&
  {Brooks}}{{Ricci} et~al.}{2010}]{Ricci2010}
{Ricci} L.,  {Testi} L.,  {Natta} A.,   {Brooks} K.~J.,  2010, \mn@doi [\aap]
  {10.1051/0004-6361/201015039},
  \href{http://ukads.nottingham.ac.uk/abs/2010A%26A...521A..66R}{521, A66}

\bibitem[\protect\citeauthoryear{{Rice}, {Lodato}, {Pringle}, {Armitage}  \&
  {Bonnell}}{{Rice} et~al.}{2004}]{Rice2004}
{Rice} W.~K.~M.,  {Lodato} G.,  {Pringle} J.~E.,  {Armitage} P.~J.,   {Bonnell}
  I.~A.,  2004, \mn@doi [\mnras] {10.1111/j.1365-2966.2004.08339.x},
  \href{http://adsabs.harvard.edu/abs/2004MNRAS.355..543R}{355, 543}

\bibitem[\protect\citeauthoryear{{Rice}, {Lodato}, {Pringle}, {Armitage}  \&
  {Bonnell}}{{Rice} et~al.}{2006}]{Rice2006}
{Rice} W.~K.~M.,  {Lodato} G.,  {Pringle} J.~E.,  {Armitage} P.~J.,   {Bonnell}
  I.~A.,  2006, \mn@doi [\mnras] {10.1111/j.1745-3933.2006.00215.x},
  \href{http://ukads.nottingham.ac.uk/abs/2006MNRAS.372L...9R}{372, L9}

\bibitem[\protect\citeauthoryear{{Rosswog}}{{Rosswog}}{2015}]{Rosswog2015}
{Rosswog} S.,  2015, \mn@doi [\mnras] {10.1093/mnras/stv225},
  \href{http://ukads.nottingham.ac.uk/abs/2015MNRAS.448.3628R}{448, 3628}

\bibitem[\protect\citeauthoryear{{Shi} \& {Chiang}}{{Shi} \&
  {Chiang}}{2014}]{Shi2014}
{Shi} J.-M.,  {Chiang} E.,  2014, \mn@doi [\apj] {10.1088/0004-637X/789/1/34},
  \href{http://ukads.nottingham.ac.uk/abs/2014ApJ...789...34S}{789, 34}

\bibitem[\protect\citeauthoryear{{Springel}}{{Springel}}{2005}]{Springel2005}
{Springel} V.,  2005, \mn@doi [\mnras] {10.1111/j.1365-2966.2005.09655.x},
  \href{http://adsabs.harvard.edu/abs/2005MNRAS.364.1105S}{364, 1105}

\bibitem[\protect\citeauthoryear{{Stewart} \& {Leinhardt}}{{Stewart} \&
  {Leinhardt}}{2009}]{Stewart2009}
{Stewart} S.~T.,  {Leinhardt} Z.~M.,  2009, \mn@doi [\apjl]
  {10.1088/0004-637X/691/2/L133},
  \href{http://ukads.nottingham.ac.uk/abs/2009ApJ...691L.133S}{691, L133}

\bibitem[\protect\citeauthoryear{{V\"{o}lk}, {Jones}, {Morfill}  \&
  {Roeser}}{{V\"{o}lk} et~al.}{1980}]{Volk1980}
{V\"{o}lk} H.~J.,  {Jones} F.~C.,  {Morfill} G.~E.,   {Roeser} S.,  1980, \aap,
  \href{http://ukads.nottingham.ac.uk/abs/1980A&A....85..316V}{85, 316}

\bibitem[\protect\citeauthoryear{{Wada}, {Tanaka}, {Suyama}, {Kimura}  \&
  {Yamamoto}}{{Wada} et~al.}{2009}]{Wada2009}
{Wada} K.,  {Tanaka} H.,  {Suyama} T.,  {Kimura} H.,   {Yamamoto} T.,  2009,
  \mn@doi [\apj] {10.1088/0004-637X/702/2/1490},
  \href{http://ukads.nottingham.ac.uk/abs/2009ApJ...702.1490W}{702, 1490}

\bibitem[\protect\citeauthoryear{{Walmswell}, {Clarke}  \&
  {Cossins}}{{Walmswell} et~al.}{2013}]{Walmswell2013}
{Walmswell} J.,  {Clarke} C.,   {Cossins} P.,  2013, \mn@doi [\mnras]
  {10.1093/mnras/stt314},
  \href{http://ukads.nottingham.ac.uk/abs/2013MNRAS.431.1903W}{431, 1903}

\bibitem[\protect\citeauthoryear{{Weidenschilling}}{{Weidenschilling}}{1977}]{%
Weidenschilling1977}
{Weidenschilling} S.~J.,  1977, \mnras,
  \href{http://ukads.nottingham.ac.uk/abs/1977MNRAS.180...57W}{180, 57}

\bibitem[\protect\citeauthoryear{{Whipple}}{{Whipple}}{1972}]{Whipple1972}
{Whipple} F.~L.,  1972, in {Elvius} A.,  ed., From Plasma to Planet. p.~211

\bibitem[\protect\citeauthoryear{{Wilkinson}, {Mehlig}  \&
  {Bezuglyy}}{{Wilkinson} et~al.}{2006}]{Wilkinson2006}
{Wilkinson} M.,  {Mehlig} B.,   {Bezuglyy} V.,  2006, \mn@doi [Physical Review
  Letters] {10.1103/PhysRevLett.97.048501},
  \href{http://adsabs.harvard.edu/abs/2006PhRvL..97d8501W}{97, 048501}

\bibitem[\protect\citeauthoryear{{Windmark}, {Birnstiel}, {Ormel}  \&
  {Dullemond}}{{Windmark} et~al.}{2012}]{Windmark2012}
{Windmark} F.,  {Birnstiel} T.,  {Ormel} C.~W.,   {Dullemond} C.~P.,  2012,
  \mn@doi [\aap] {10.1051/0004-6361/201220004},
  \href{http://ukads.nottingham.ac.uk/abs/2012A%26A...544L..16W}{544, L16}

\bibitem[\protect\citeauthoryear{{Youdin} \& {Goodman}}{{Youdin} \&
  {Goodman}}{2005}]{Youdin2005}
{Youdin} A.~N.,  {Goodman} J.,  2005, \mn@doi [\apj] {10.1086/426895},
  \href{http://ukads.nottingham.ac.uk/abs/2005ApJ...620..459Y}{620, 459}

\bibitem[\protect\citeauthoryear{{Youdin} \& {Lithwick}}{{Youdin} \&
  {Lithwick}}{2007}]{Youdin2007}
{Youdin} A.~N.,  {Lithwick} Y.,  2007, \mn@doi [\icarus]
  {10.1016/j.icarus.2007.07.012},
  \href{http://ukads.nottingham.ac.uk/abs/2007Icar..192..588Y}{192, 588}

\bibitem[\protect\citeauthoryear{{Young} \& {Clarke}}{{Young} \&
  {Clarke}}{2015}]{Young2015}
{Young} M.,  {Clarke} C.~J.,  2015, \mnras~submitted

\bibitem[\protect\citeauthoryear{{Zhang}, {Blake}  \& {Bergin}}{{Zhang}
  et~al.}{2015}]{Zhang2015}
{Zhang} K.,  {Blake} G.~A.,   {Bergin} E.~A.,  2015, \mn@doi [\apjl]
  {10.1088/2041-8205/806/1/L7},
  \href{http://ukads.nottingham.ac.uk/abs/2015ApJ...806L...7Z}{806, L7}

\bibitem[\protect\citeauthoryear{{Zsom}, {Ormel}, {G{\"u}ttler}, {Blum}  \&
  {Dullemond}}{{Zsom} et~al.}{2010}]{Zsom2010}
{Zsom} A.,  {Ormel} C.~W.,  {G{\"u}ttler} C.,  {Blum} J.,   {Dullemond} C.~P.,
  2010, \mn@doi [\aap] {10.1051/0004-6361/200912976},
  \href{http://ukads.nottingham.ac.uk/abs/2010A%26A...513A..57Z}{513, A57}

\makeatother
\end{thebibliography}
    \bibliographystyle{mnras_edit}
}

\bsp

\appendix
\section{Gravitational Softening in 2D}
\label{Sec:Gravity}

Since each particle in a two dimensional disc simulation represents a column of
gas with a scale height $H$, it is necessary to soften the gravitational force,
which no longer has $r^{-2}$ dependence for $r < H$. Often a Plummer potential
is used for this purpose, $\phi = (r^2 + \epsilon^2)^{-1/2}$, where 
$\epsilon \approx 0.6 H$ is often used. As shown by \citet{Muller2012}, this 
underestimates the force for $r < H$ since it produces a force with
$\propto r$ rather than $f \propto r^{-1}$. For an isothermal column with 
$\rho = \rho_0 \exp(- z^2 / (2 H^2))$, integrating over the vertical density
gives the following acceleration at the mid-plane a distance $r$ from the particle,
which gives the gravitational acceleration, $g$,
\begin{equation}
 g(r) = - \frac{G m_p}{r^2} I(\frac{r}{H}),
\end{equation}
where 
\begin{equation}
 I(x) = \sqrt{\frac{8}{\pi}} \frac{x^3}{8} \exp\left(\frac{x^2}{4}\right)
\left[ K_1\left(\frac{x^2}{4}\right) - K_0\left(\frac{x^2}{4}\right)\right],
\label{Eqn:SoftGauss}
\end{equation}
where $K_0$ and $K_1$ are modified Bessel functions of the \emph{second} kind 
\citep{Muller2012}. Since \autoref{Eqn:SoftGauss} is expensive to evaluate directly
we instead use a rational polynomial approximation, constructed to reproduce
the limits $I(x) \rightarrow 1$ as $x \rightarrow \infty$ and 
$I(x) \rightarrow \sqrt{2/\pi} x$ as  $x \rightarrow 0$. Using a polynomial of
the form 
\begin{equation}
 I_p (x) = \frac{1 + A x^{-1}}{1 + B x^{-1} + A f x^{-2}}, \label{Eqn:2DSoft1}
\end{equation}
where $f = \sqrt{\pi / 2}$, reproduces these limits. For the fitting constants,
we use $A = 3/2$, and set $B$ to interpolate through the exact value of $I$
at $r=H$, $I(1)$. This gives $B = (1 + A)/ I(1) - (1 + A f)$. For this choice of 
parameters the maximum relative error is less than 1 per cent.

\begin{figure}
 \centering
\includegraphics[width=\columnwidth]{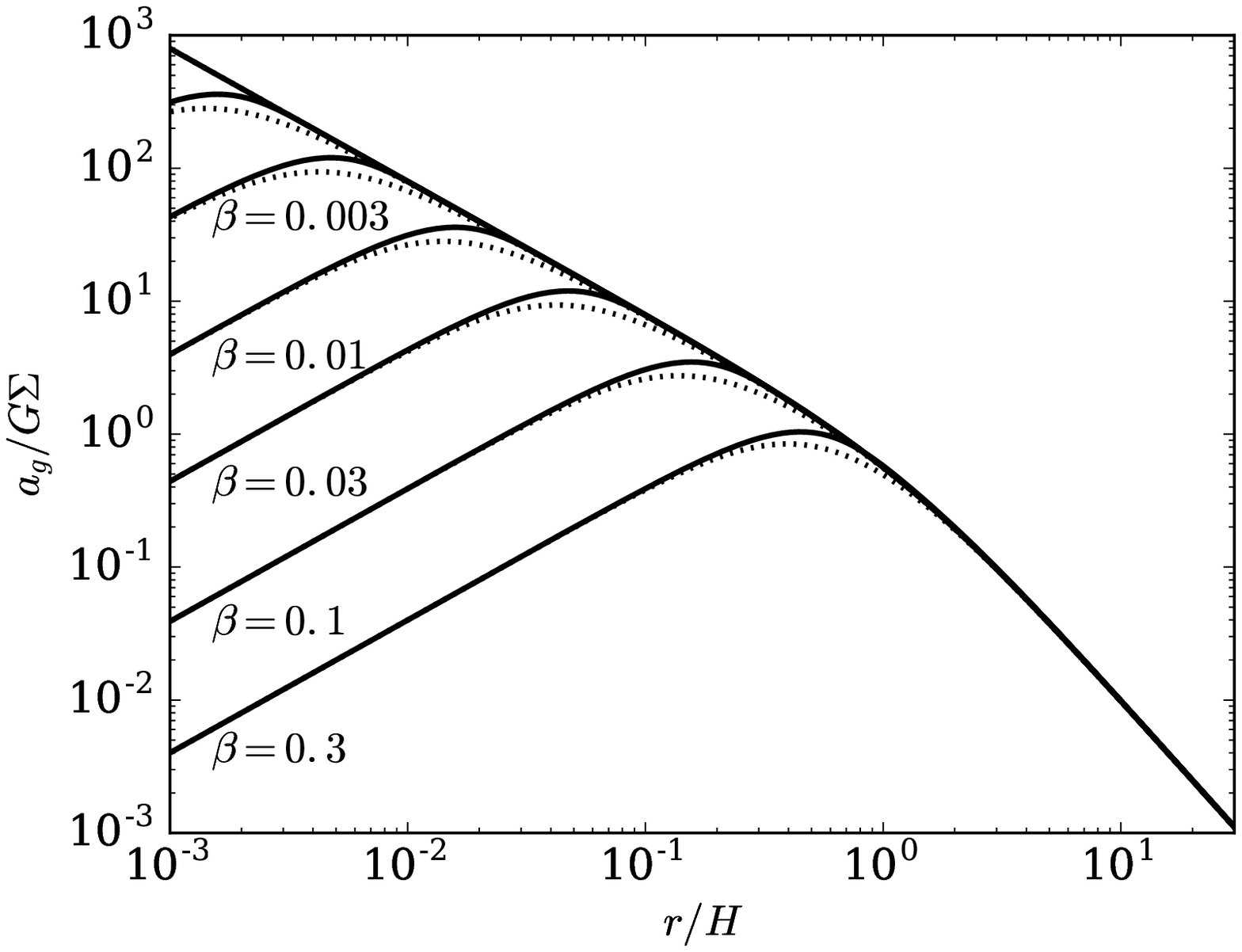}
\caption{Specific gravitational force, $a /G\Sigma = I_{\rm SPH} / r^2$ at $z=0$, from a 
particle with the density given by \autoref{Eqn:Soft2Drho} (solid lines), and the 
approximate form (\autoref{Eqn:SoftApprox} and \autoref{Eqn:chiApprox}, dotted lines)
for a range of different softening parameters. The approximation gives excellent agreement
with the exact force in the limits $r \gg \beta H$ and $r \ll \beta H$, but under estimates 
the force  by 25 per cent at $r=\beta H$.
}
\label{Fig:Soft}
\end{figure}

We note that the vertical structure in self-gravitating discs differs from
a Gaussian. \citet{Muller2012} found a only 10 per cent difference between
the force calculated assuming a Gaussian and isothermal density slab density
profiles. The actual force error introduced by using a vertically averaged
must be smaller than this, since for a disc with a Toomre $Q$ parameter of 
$Q = 1$ both the central star and disc self-gravity have equal importance
\citep{Lodato2007}. 

Since the vertically integrated force diverges as $r^{-1}$ for small separations
it is necessary to further soften the force between neighbouring particles. Usually
this is done by assuming that the mass of each particle is distributed according
to the SPH kernel, $\rho_i(r) = m_i W(r, h_i)$, which leads to a self-consistent
definition of the SPH density estimate and a softening for the potential
that can be constructed using the Poisson equation \citep{Dehnen2001,Price2007}.
By a similar analogy a particle density for disc simulations can be written as 
\begin{equation}
 \rho_{{\rm 2D},i}(R, z) = m_i W_g(R, \epsilon_i) 
\frac{\exp\left(-\tfrac{z^2}{2 H_i^2}\right)}{\sqrt{2 \pi} H_i}.
\end{equation}
Integrating $\rho_{{\rm 2D},i}$ over $z$ gives the usual definition for the 
(surface) density in two dimensions when $\epsilon_i = h_i$. 
The idea now is to solve the Poisson equation
for this density distribution and use it to construct a softening function
$I_{\rm SPH}$. For simplicity, we use a 2D Gaussian kernel rather than a Wendland
or cubic spline kernel for computing the softening, i.e.
\begin{equation}
 \rho_{{\rm 2D},i}(R, z) = \frac{m_i}{(2 \pi)^{3/2} \epsilon_i^2 H_i}
\exp\left[-\left(\frac{z^2}{2 H_i^2}+\frac{R^2}{2 \epsilon_i^2}\right)\right].
\label{Eqn:Soft2Drho}
\end{equation}
For $r \ll \epsilon, H$; the density is approximately constant so we can expect
$I_{\rm SPH} \propto x^3$. For $\epsilon = H$ the density is spherically 
symmetric, and $I_{\rm SPH} = \sqrt{2/\pi} x^3 / 3$ for $x \ll H$. In this 
way, we can expect a function of the form 
\begin{equation}
I_{\rm SPH}(x) \approx \left[I_p(x)^{-1} + 
               \left(\sqrt{\frac{2}{\pi}} \frac{x^3}{3} \chi\right)^{-1}\right]^{-1},
\label{Eqn:SoftApprox}
\end{equation}
to be reasonably accurate. The factor $\chi \equiv \chi(\epsilon, H)$
takes into account the deviation from spherical symmetry. To find an approximation
for $\chi(\epsilon,H)$, we solve for the force at the disc mid-plane by, taking the
Fourier transform of the Poisson equation, and solving when $z = 0$, giving
\begin{equation}
 I_{\rm SPH}(x) = \frac{x^3}{2 (2\pi)^{3/2}} 
\int_0^\pi \frac{I_0(\lambda(\theta)) - I_1(\lambda(\theta))}
		{[(1 - \beta^2)\cos^2\theta + \beta^2]^{3/2}}
	  \sin^3\theta\,{\rm d}\theta,
\label{Eqn:Soft2De}
\end{equation}
where $\beta = \epsilon / H$,  $I_0$ and $I_1$ are modified Bessel functions of the
\emph{first} kind and
\begin{equation}
 \lambda(\theta) = \frac{x^2 \sin^2\theta}{4((1 - \beta^2)\cos^2\theta + \beta^2)}.
\end{equation}
 For $R \gg \epsilon, H$, and $\epsilon \ll r \ll H$, we have 
$I_{\rm SPH}(x) \rightarrow 1$ and $I(x) \sim \sqrt{2/\pi} x$, as before. For
$r \ll \epsilon$; $\lambda \ll 1$ and \autoref{Eqn:Soft2De} can be integrated,
providing an expression for $\chi$, 
\[\chi(\beta) = \frac{3}{2}
 \begin{cases}
  \frac{1}{\beta^2-\beta^4} - \log\left(\frac{1+\sqrt{1-\beta^2}}{\beta}\right)(1-\beta^2)^{-3/2} & \beta < 1, \\
  \frac{2}{3} & \beta = 1, \\
  \frac{1}{\beta^2-\beta^4} + \sin^{-1}\left(\frac{\sqrt{\beta^2-1}}{\beta}\right)(\beta^2-1)^{-3/2} & \beta > 1. 
 \end{cases}
\]
Once again we use a simple rational polynomial approximation for $\chi$ to speed up the
force evaluation,
\begin{equation}
 \chi(\beta) \approx \frac{3}{2\beta^2}
                     \frac{1 + \pi \beta}{1 + \tfrac{3}{2}(\pi -1) \beta + 2 \beta^2},
\label{Eqn:chiApprox}
\end{equation}
which is accurate to 0.1 per cent and reproduces the correct limits for $\beta \rightarrow 0$,
$\beta =1$ and $\beta \rightarrow \infty$. \autoref{Eqn:SoftApprox} and \autoref{Eqn:chiApprox}
form the basis of the 2D softening potential that we use.

The exact force from the density function $\rho_{{\rm 2D},i}$ and the approximation are shown
in \autoref{Fig:Soft}. While the approximate form underestimates the force by 25 per cent
close to the softening length, the agreement is excellent at both larger and smaller 
separations.

\label{lastpage}
\end{document}